\documentclass[aps,reprint,twocolumn,showpacs,preprintnumbers,amsmath,amssymb,nofootinbib,superscriptaddress,showkeys]{revtex4-1}

\usepackage{bm}
\usepackage{epsfig}
\usepackage{slashed}

\begin{document}

\title{Power Counting and
Perturbative One Pion Exchange in Heavy Meson Molecules}
 \author{M. Pav\'on Valderrama}\email{m.pavon.valderrama@ific.uv.es}
 \affiliation{Instituto de F\'{\i}sica Corpuscular (IFIC), Centro Mixto CSIC-Universidad de Valencia, Institutos de Investigaci\'on de Paterna, Aptd. 22085, E-46071 Valencia, Spain}

\date{\today}

\begin{abstract}
\rule{0ex}{3ex}
We discuss the possible power counting schemes that can be applied
in the effective field theory description of heavy meson molecules,
such as the $X(3872)$ or the recently discovered $Z_b(10610)$
and $Z_b(10650)$ states.
We argue that the effect of coupled channels is suppressed
by at least two orders in the effective field theory expansion,
meaning that they can be safely ignored at lowest order.
The role of the one pion exchange potential between the heavy mesons,
and in particular the tensor force, is also analyzed. 
By using techniques developed in atomic physics for handling power-law
singular potentials, which have been also successfully employed
in nuclear physics, we determine the range of center-of-mass
momenta for which the tensor piece of the one pion exchange
potential is perturbative.
In this momentum range, the one pion exchange potential can be considered
a subleading order correction, leaving at lowest order a very simple
effective field theory consisting only on contact-range interactions.
\end{abstract}

\pacs{13.75.Lb,14.40.Lb,14.40.Nd,14.40Pq,14.40Rt}

\maketitle

\section{Introduction}

Heavy meson molecules are a long established theoretical prediction
of hadronic physics~\cite{Voloshin:1976ap,DeRujula:1976qd,Tornqvist:1991ks,
Tornqvist:1993ng,Manohar:1992nd,Ericson:1993wy}.
The discovery of the $X(3872)$ by the Belle collaboration~\cite{Choi:2003ue},
and the subsequent confirmation by CDF~\cite{Acosta:2003zx},
D0~\cite{Abazov:2004kp} and BABAR~\cite{Aubert:2004ns},
has provided so far the strongest candidate
for a bound state of heavy mesons.
Owing to the closeness of the $X(3872)$ to the $D^{0*}\bar{D}^0$ threshold,
an interpretation in terms of a bound state of
these two mesons is both appealing and natural.
In turn,
the recent finding of the $Z_b(10610)$ and $Z_b(10650)$ resonances
by Belle~\cite{Collaboration:2011gj,Belle:2011aa},
which are located just a few ${\rm MeV}$ above the $B^*\bar{B}$
and $B^*\bar{B}^*$ thresholds respectively,
also calls for a molecular description.
Moreover, from heavy quark spin symmetry~\cite{Isgur:1989vq,Isgur:1989ed}
we should expect other low-lying states, the partners of the $Z_b$'s,
in the bottom sector~\cite{Voloshin:2011qa,Mehen:2011yh}.

The shallow nature of the aforementioned candidates for molecular states,
apart from being instrumental in their identification,
also indicates the existence of a separation of
scales between long and short range dynamics.
The heavy mesons are far apart from each other
and consequently are not able to resolve the details
of the short range interaction that may be ultimately responsible
for binding them.
We therefore expect that heavy meson molecules will be amenable
to an effective field theory (EFT) description of
their properties and decays~\cite{Braaten:2003he,AlFiky:2005jd,Fleming:2007rp}.

In the EFT formalism, the long range interaction between
two heavy mesons (composed of a heavy and a light quark)
is constrained by the low-energy symmetries
of the system,
in particular chiral symmetry,
which determines the pion-exchange dynamics.
In turn, the short range interaction
is mimicked by local, contact-range operators or counterterms.
A second ingredient of the EFT description is
the existence of a power counting, that is,
an organizational principle
from which operators can be ordered from more to less relevant.
From power counting we expect to be able to express operators and observables
as a power series in term of a small expansion parameter $x_0$
\begin{eqnarray}
{O}_{\rm EFT} = \sum_{\nu = \nu_0}^{\nu_{\rm max}} \hat{O}^{(\nu)} x_0^{\nu} + 
\mathcal{O}(x_0^{\nu_{\rm max} + 1}) \, ,
\end{eqnarray}
where $\nu_0$ is the order at which the EFT expansion starts,
$\nu_{\rm max}$ the order at which we perform the calculation,
and $x_0$, the expansion parameter, can be written as the ratio
\begin{eqnarray}
x_0 = \frac{Q}{\Lambda_0} \, ,
\end{eqnarray}
with $Q$ ($\Lambda_0$) the generic energy scale associated
with the long (short) range physics.
A priori power counting entails a remarkable advantage:
the relative error of a calculation of order $\nu$ is known to be $x_0^{\nu+1}$.
Of course, the realization of this promise of accuracy depends on the
implementation details and on the unambiguous identification of
the expansion parameter.
This manuscript will try to deal with this problem.

At this point a natural question arises: how important are the pion exchanges
in the description of the heavy meson molecules?
Of course, having a bound state requires the non-perturbative
treatment of a certain subset of the meson-antimeson interaction,
but not necessarily the one-pion exchange (OPE) potential.
A nice illustration is provided by the $X(3872)$,
in which the distance between the heavy meson and antimeson is so large
($\sqrt{\langle r^2 \rangle} \sim 10\,{\rm fm}$)
that even pions may not be clearly
distinguished.
This means that we can use a contact-range, pionless EFT
in which pions are already considered to be short-ranged~\cite{vanKolck:1998bw,Chen:1999tn}.
In such a case we are left with a much simpler theory.
Even though the hard scale is the pion mass, $\Lambda_0 = m_{\pi}$,
which is an extremely light scale in hadronic physics,
the soft scale $Q$ can be even lighter
if the molecular state is close enough to threshold.
In particular, $Q$ can be identified with the wave number of the bound state,
that is, $Q = \sqrt{2 \mu_R B}$, where $\mu_R$ is the reduced mass of
the two particle system and $B$ the binding energy.
For the $D^{*0} \bar{D}^0$ mesons conforming the $X(3872)$,
$Q$ is of the order of mere tens of ${\rm MeV}$.
Thus the EFT expansion is expected to converge fairly well,
providing a motivation and explaining the success
of contact-range descriptions of the $X(3872)$~\cite{Close:2003sg,
Voloshin:2003nt,Braaten:2003he,Voloshin:2005rt}.

However, the applicability of pionless EFT is subjected to limitations.
In the $X(3872)$ we can list two: on the one hand,
the pion exchanged between the $D^{*0}$ and $\bar{D}^0$ meson
is almost on the mass shell, and consequently
its effects spread over a much larger range
than expected. 
That is, the irrelevance of pion exchanges is not so evident
as it appears at first sight.
Luckily, as shown in X-EFT~\cite{Fleming:2007rp},
pions are actually perturbative in the $X(3872)$.
On the other,
if we consider the charged $D^{*+} D^{-}$ component of the $X(3872)$,
which is essential for explaining certain branching
ratios~\cite{Gamermann:2009fv,Gamermann:2009uq},
the associated soft scale is $Q \sim 125\,{\rm MeV}$,
of the order of $m_{\pi}$.
Thus the explicit inclusion of the charged channel
may lie outside the range of applicability of a pionless EFT.
If we now consider the bottom sector,
the $Z_b(10610)$ and $Z_b(10650)$ resonances are a only few ${\rm MeV}$
away from the $B^*\bar{B}$ and $B^*\bar{B}^*$ threshold~\footnote{
It is interesting to notice that even though the original experimental
analysis by Belle~\cite{Collaboration:2011gj,Belle:2011aa} locates
the two states above the $B^*\bar{B}$ and $B^*\bar{B}^*$
thresholds, this depends on the Breit-Wigner
parametrization employed for the $Z_b$'s 
as stressed in Ref.~\cite{Cleven:2011gp}.
This work also indicates that the two $Z_b$'s may be located below threshold
and hence be bound states (instead of two-particle resonances) after all.}.
However, the large reduced mass of these systems imply that $Q \sim m_{\pi}$
corresponds to a binding energy of merely $4\,{\rm MeV}$
at which the wave function will start to probe the pions,
requiring a non-perturbative theory~\cite{Nieves:2011zz}.

In all the previous examples the inclusion of pion exchanges may be required
for a proper EFT description of the molecular states.
This can be rather cumbersome, owing to the rich angular momentum
coupled channel structure triggered by tensor forces,
especially in the $D^*\bar{D}^*$ and $B^*\bar{B}^*$ cases.
However, there exists a binding energy range in which the OPE potential
is subleading with respect to the contact range interactions
and hence perturbative.
In this energy window the pionfull EFT will still be a contact-range theory
at the lowest (or leading) order (${\rm LO}$),
as pion exchanges will not enter until next-to-leading order (${\rm NLO}$).
In the two-nucleon system the paradigmatic example of this kind of EFT
is the Kaplan, Savage and Wise (KSW)
counting~\cite{Kaplan:1998tg,Kaplan:1998we},
which served as inspiration for X-EFT~\cite{Fleming:2007rp}.

The question we want to answer is:
where does the limit between perturbative and non-perturbative pions stand?
If we take a second look to the deuteron in the two nucleon system,
which shares many similarities with the heavy meson molecules,
we see that the convergence of the EFT with perturbative pions
is numerically marginal.
As shown by Fleming, Stewart and Mehen (FMS)~\cite{Fleming:1999ee}
by a thorough next-to-next-to-leading order (${\rm N^2LO}$) calculation,
the convergence of the KSW counting in the deuteron case is limited
at most to $\Lambda_0 \sim 100\,{\rm MeV}$, a rather low figure
(indeed smaller than $m_{\pi}$).
Taking into account that in the deuteron the wave number
is $\gamma = 45\,{\rm MeV}$,
the previous breakdown scale translates into a rather slow convergence rate.
In fact, the theory would not converge at all had the deuteron
be bound by about $4\,{\rm MeV}$ or more.
The purpose of this work is therefore to find the corresponding
$B_{\rm max}$ below which pions are perturbative, as this limit
will gives us essential information about the convergence of
the EFT for heavy meson molecules.

Of course, performing a ${\rm N^2LO}$ calculation is beyond
the scope of the present paper.
Apart from that, the scarce experimental input available
on heavy meson-antimeson systems makes the previous task
impractical: we do not have the information required to
fix the counterterms.
Therefore, we need to resort to a more indirect path.
In the two-nucleon system the solution was provided
by Birse~\cite{Birse:2005um}.
The idea is to consider the OPE in the chiral limit, in which it reduces
to a pure power-law potential of the type $1/r^3$.
The interesting thing here is that these kind of potentials have been
studied and analyzed in detail in the field of atomic physics.
The long-range solutions of OPE in the chiral limit
can be expressed in terms of a a particular type of series,
as shown by Cavagnero~\cite{Cavagnero:1994zz} and Gao~\cite{Gao:1999zz}.
Birse was able to show that one can extract the breakdown scale of
a theory containing perturbative pions with the help of these techniques,
and the results help to explain remarkably well the
lengthier ${\rm N^2LO}$ calculations of FMS~\cite{Fleming:1999ee}.
In this work we will extend the observations of Birse
to the peculiarities of the tensor force
in heavy meson molecules.

The article is structure as follows: in Sect.~\ref{sec:gen}
we present a brief overview of certain key EFT ideas
that will help to put the results of this work
into proper context.
In Sect.~\ref{sec:LO}
we will write the OPE and contact range potentials
between a heavy meson and antimeson,
and in Sect.~\ref{sec:pc} we will discuss the different power counting
schemes that we can apply.
In Sect.~\ref{sec:pert} we will extract the breakdown scale of
perturbative pion theories, and finally,
in Sect.~\ref{sec:dis} we will discuss the results and their implications
on the EFT treatment of heavy molecular states.
We have also included an Appendix containing the technical details involved
in the derivation of the EFT potential at lowest order.

\section{General Considerations}
\label{sec:gen}

The purpose of this section is to provide a quick review of
the EFT formalism for non-relativistic two-body systems.
The discussion is heavily based on the EFT formulation of
the two-nucleon system (see Refs.~\cite{Beane:2000fx,Bedaque:2002mn,Epelbaum:2005pn,Epelbaum:2008ga,Machleidt:2011zz} for reviews),
which can be trivially translated
and applied to heavy meson molecules with minor modifications,
as demonstrated in Ref.~\cite{Nieves:2011zz}.
Thus, we begin with naive dimensional power counting,
as originally proposed by Weinberg~\cite{Weinberg:1990rz,Weinberg:1991um},
and then explain the modifications that have been required to
successfully formulate a non-relativistic EFT to two-body
systems forming shallow bound states~\cite{Kaplan:1998tg,Kaplan:1998we,
Gegelia:1998gn,Birse:1998dk,vanKolck:1998bw}.
As we are dealing with heavy meson molecules, we will include
notation related to heavy quark symmetry.
In what follows $m_Q$ is the mass of the heavy quark conforming a heavy meson,
and ${\rm H} = {\rm P, P^*}$ is used to denote a generic heavy meson
with orbital angular momentum $l=0$ between the heavy quark
and the light quark.
The heavy meson $\rm P$ ($\rm P^*$) has total spin $s =0$ ($s=1$)
and hence it is a pseudoscalar (vector) meson.
If we are specifically dealing with the charm or bottom sector we will
particularize the $\rm P$, $\rm P^*$ heavy meson notation
by $\rm D$, $\rm D^*$ and $\rm B$, $\rm B^*$.
We also use $\rm D^{(*)}$ and $\rm B^{(*)}$ as a generic
for the pseudoscalar/vector cases.
We will only consider the case in which the light quark is
the $u$ or $d$ quark.
The extension to the strange sector is straightforward.

\subsection{Power Counting}

The formulation of EFT depends on the existence of a separation of scales:
we can distinguish between $Q$, the low energy scale that characterizes
the physics we are interested in, and $\Lambda_0$,
the high energy scale at which the effective description
we are using stops to be applicable.
In the EFT framework, the two-body potential can be expanded
as a power series on the ratio of these scales, leading to
\begin{eqnarray}
V_{\rm EFT} = \sum_{\nu = \nu_0}^{\nu_{\rm max}} V^{(\nu)}
+ \mathcal{O} \left[
{\left( \frac{Q}{\Lambda_0} \right)}^{\nu_{\rm max} + 1} \right] \, ,
\end{eqnarray}
where $\nu_0 \geq -1$ is the order at which the expansion begins and
$\nu_{\rm max}$ the order at which we perform the calculation.
In chiral (nuclear~\cite{Weinberg:1990rz,Weinberg:1991um}
and heavy hadron~\cite{Wise:1992hn})
EFT the generic scale $Q$ usually
includes the momenta of the two interacting particles and
the mass of the pion, while $\Lambda_0$ refers to the mass of
the rho meson or the momentum scale at which the internal
structure of the two particles starts to be resolved.

As can be seen,
we expect the theoretical error in the determination of the EFT potential
to decrease, at least as long as $Q \leq \Lambda_0$.
At this point we should take into account that the light scales include
on the one hand the pion mass $m_{\pi}$, which does not change~\footnote{
Unless we are considering chiral extrapolations.},
and on the other the momenta $Q \sim p, p'$ of the two heavy mesons,
which can vary.
This means that the expansion of the potential
(and the scattering observables) is only valid
for sufficiently small $p,p'$.
To avoid the related divergences with taking $p, p' \geq \Lambda_0$ in loops
we usually include a cut-off $\Lambda$ that serves as an intermediate
scale between $Q$ and $\Lambda_0$
(i.e. we take $Q \leq \Lambda \leq \Lambda_0$).

\subsection{The Scaling of Operators}

The power counting assignment of a certain operator tells us
about its scaling properties. 
A contribution to the (momentum space) potential is assigned
the order $\nu$ if it scales as
\begin{eqnarray} 
V^{(\nu)}(\lambda\,Q) = \lambda^{\nu}\,V^{(\nu)}(Q) \, ,
\end{eqnarray}
under the rescaling of all the light scales by a factor $\lambda$.
This means that if we reduce the momentum or the pion mass by a factor
$R = \frac{1}{\lambda}$, with $R \leq 1$,
thus increasing the separation of scales
by a factor of $\lambda = \frac{1}{R} \geq 1$,
the size of the order $\nu$ contributions will decrease as $R^{\nu}$.
That is, such a contribution becomes smaller
the larger the scale separation,
justifying their power counting assignment.

Scaling is very interesting in the sense that it determines
the behaviour of the potential in coordinate space.
If we consider the order $\nu$ contribution potential in momentum space,
we simply have~\footnote{
Notice that we have restricted ourselves to the local potential case.
Non-localities only appear at high orders in the chiral expansion
($\nu = 4$ in the two-nucleon case) and in addition they are
suppressed by the mass of the heavy mesons.
}
\begin{eqnarray} 
V^{(\nu)}(\lambda\,\vec{q}, \lambda\,Q) =
\lambda^{\nu} \, V^{(\nu)}(\vec{q}, Q) \, ,
\end{eqnarray}
where we have now explicitly considered the dependence on the momentum
exchanges between the two particles, $\vec{q} = \vec{p} - \vec{p}\,'$.
After Fourier-transforming into coordinate space,
the previous scaling translates to
\begin{eqnarray}
V^{(\nu)}(\frac{\vec{r}}{\lambda}, \lambda\,Q) = \lambda^{3+\nu}\,
V^{(\nu)}(\vec{r},Q) \, ,
\end{eqnarray}
which admits two kind of general solutions, contact range and finite range.
The contact range solution is trivial to construct from
the Dirac $\delta$-function and its derivatives,
yielding
\begin{eqnarray}
V_{\rm C}^{(\nu)}(\vec{r},Q) = C_{\nu} \, \partial^{\nu} \delta(\vec{r}) \, , 
\end{eqnarray}
where $\partial$ denotes a general derivative of the Dirac $\delta$.
Of course, parity constraints imply that this kind of contribution
only appears at $\nu = 2 n$.
On the other hand, the finite range solution must comply to the form
\begin{eqnarray}
V^{(\nu)}(\vec{r},Q) = \frac{F_{\nu}}{r^{3+\nu}}\,f^{(\nu)}(Q \, r) \, ,
\end{eqnarray}
with $f(x)$ an arbitrary (non exclusively power-law) function,
which decays exponentially at large distances
as it stems from meson exchanges.

In the previous equations $C_{\nu}$ and $F_{\nu}$ are constants with
dimensions of $1 / \left[ {\rm energy} \right]^{\nu+2}$.
From power counting we expect the related energy scale to be $\Lambda_0$,
that is, 
\begin{eqnarray}
\label{eq:strength}
C_{\nu}\sim \frac{1}{\Lambda_0^{\nu+2}}
\quad\mbox{and}\quad
F_{\nu}\sim \frac{1}{\Lambda_0^{\nu+2}}
\, .
\end{eqnarray}
However, as we will see in the following paragraphs,
sometimes there is an unexpected contamination
of $C_{\nu}$ and $F_{\nu}$ by a light scale.
This will trigger a change in the power counting of the potential,
promoting a certain contribution from a higher to a lower order.
An illustration is given by the scaling arguments discussed
in the introduction of Ref.~\cite{PavonValderrama:2007nu}.

\subsection{Iteration}

If we are considering scattering states,
we should plug the potential into the Lippmann-Schwinger equation,
\begin{eqnarray}
T = V + V G_0 T \, ,
\end{eqnarray}
where $V$ is the EFT potential, $T$ is the T-matrix and
$G_0 = 1 / (E - H_0)$ the resolvent operator.
Analogously, if we are considering heavy meson molecules
we should iterate the potential in the bound state equation
\begin{eqnarray}
| \Psi_B \rangle = G_0 V \, | \Psi_B \rangle \, ,
\end{eqnarray}
to obtain the binding energy and the wave function $| \Psi_B \rangle$.

Within the EFT framework the previous equations are expected to be re-expanded
according to the power counting of the potential.
In this way, we are guaranteed to be able to estimate the error properly. 
For this, we need to take into account that the scaling of the resolvent
operator is given by
\begin{eqnarray}
G_0 \sim \mu_R\,Q \, ,
\end{eqnarray}
where $\mu_R$ is the reduced mass of the two body system,
which for a non-relativistic theory can be considered
a high energy scale, $\mu_R \sim \Lambda_0$.
From the point of view of power counting, the scaling of $G_0$ means
that only contributions to the EFT potential such that
$\mathcal{O}(G_0 V) \sim 1$ should be iterated,
i.e. the order $Q^{-1}$ piece of the potential.

The existence of shallow bound states between heavy mesons imply that
there is a contribution to the potential of order $Q^{-1}$.
However, when we compute the EFT potential from the chiral lagrangian
we only find contributions of order $Q^0$ and higher.
The OPE potential is order $Q^0$, as can be trivially checked from
\begin{eqnarray}
V_{\rm OPE}(\vec{q}) \propto \frac{q^2}{q^2 + m_{\pi}^2} \, ,
\end{eqnarray}
and it is also clear that contact range potentials
must be at least of order $Q^0$.
This means that naive dimensional analysis is not enough to determine
accurately the order of an operator if there is a bound state.
What is missing is the unexpected contribution
from a light energy scale, as has been discussed extensively
in nuclear EFT~\cite{Kaplan:1998tg,Kaplan:1998we,Gegelia:1998gn,Birse:1998dk,vanKolck:1998bw}.
A way to see this is to consider the coupling $C_0$ of the lowest order
contact operator:
if there is a bound state, $C_0$ cannot be perturbative, that is, small.
We can account for the fact that $C_0$ is big by multiplying
the naive expectation for its size by a big number,
say $\Lambda_0 / Q$, yielding
\begin{eqnarray}
C_0(Q) \sim \frac{1}{Q\,\Lambda_0} \, ,
\end{eqnarray}
from which it is obvious that $C_0(Q)$ is of order $Q^{-1}$,
promoting the lowest order contact range potential
from order $Q^0$ to $Q^{-1}$.
A different way is to solve the bound state equation with a contact operator,
in which case we obtain that $C_0$ should scale as
\begin{eqnarray}
C_0(Q) \sim \frac{1}{\gamma\,\Lambda_0} \, ,
\end{eqnarray}
with $\gamma = \sqrt{2\mu_R\,B}$ the wave number of the heavy meson molecule,
and $B$ the binding energy.
For a shallow bound state we have $\gamma \sim Q$
and the wave number is the light scale contaminating $C_0$.

Moreover, there is a second mechanism that can lead to the promotion of
the potential, in particular OPE, to order $Q^{-1}$.
It is based on a well-known argument of Ericson and Karl~\cite{Ericson:1993wy}
about the relative strength of OPE in heavy meson molecules.
The idea is that the intrinsic strength of the OPE potential,
i.e. $F_0$ as defined in Eq.~(\ref{eq:strength}),
is roughly independent on the heavy quark mass~\footnote{
Actually, it is proportional to the axial pion coupling $g^2$,
which contains a contribution of order $m_Q^0$ plus corrections
of order $m_Q^{-1}$, $m_Q^{-2}$ and so on.}.
As a consequence, the ratio
\begin{eqnarray}
\frac{V^{(0)} G_0 V^{(0)}}{V^{(0)}}
\sim \frac{\mu_R}{\Lambda_0}\,\frac{Q}{\Lambda_0} \, ,
\end{eqnarray}
is only small for the naive identification $\mu_R \sim \Lambda_0$,
for which the ratio above scales as expected if the OPE potential
is a $Q^0$ operator.
However, while the chiral hard scale $\Lambda_0$ is fixed,
the reduced mass of the two heavy meson systems scales as $m_Q$
in the heavy quark limit.
Eventually we can have $\mu_R \gg \Lambda_0$, enhancing the pion loops
to the point of making the previous ratio to be of order $Q^0$.
In such a case the OPE potential will become non-perturbative,
as expected from Ref.~\cite{Ericson:1993wy}.

\subsection{Coupled Channels}

If we consider heavy meson molecules,
we expect from heavy quark symmetry that the energy gap between the thresholds 
of the three different combinations of pseudoscalar and vector mesons,
that is
\begin{eqnarray}
{\rm HH} = {\rm PP, PP^*, P^*P^*} \, ,
\end{eqnarray}
will shrink as $\Delta_Q \sim 1/m_Q$~\cite{Neubert:1993mb,Manohar:2000dt}.
This indicates, at first sight, the necessity of a coupled channel approach.
In such a case, the Lippmann-Schwinger equation can be written as
\begin{eqnarray}
T_{AB} = V_{AB} + V_{AC} \,G_{0}\, T_{CB} \, ,
\end{eqnarray}
where $A,B,C = {\rm PP, PP^*, P^*P^*}$.
If we are considering the scattering in the vicinity of a given threshold
$\alpha$, then we set $A = B = \alpha$ in the Lippmann-Schwinger equation.
In addition, we set the center-of-mass energy to zero $E = 0$
at the $\alpha$ threshold.
This means that we must include a proper energy shift for
the resolvent operator if $G_0$ is acting on a channel $C \neq \alpha$,
that is
\begin{eqnarray}
G_{0}^{-1}(E)\,|\,\vec{q}, C \rangle =
(E - \frac{q^2}{2\mu_C} - \Delta^{\alpha}_{C})\,|\,\vec{q}, C \rangle \, ,
\end{eqnarray}
where we have considered the inverse of the $G_0$ operator for simplicity.
In the expression above, $\mu_C$ is the reduced mass of channel $C$ and
the energy shift is given by $\Delta^{\alpha}_C = M_{\alpha} - M_C$,
with $M_{\alpha}$ and $M_C$ the total mass of channels
$\alpha$ and $C$ respectively.

However, the counting of the $G_{0}$ operator is not necessarily $Q$
on a channel $C \neq \alpha$.
If we consider the ratio of the $G_0$ operator evaluated
at the $C$ and $\alpha$ channels respectively, we find that
\begin{eqnarray}
\frac{G_{0,C}^{\alpha}(E)}{G_{0,\alpha}^{\alpha}(E)} =
\frac{2\mu_A\,E - q^2}{2\mu_C\,(E - \Delta^{\alpha}_{C}) - q^2} \sim
\left(\frac{Q}{\Lambda_C}\right)^2 \, ,
\end{eqnarray}
where $Q \sim \sqrt{2 \mu_R E} \sim q$ (we are assuming similar reduced masses),
and $\Lambda_C = \sqrt{2 \mu_C \Delta_C^{\alpha}}$ is the momentum scale
related to the coupled channel effects.
We can distinguish here two possible situations:
(i) $\Lambda_C \ll \Lambda_0$ and (ii) $\Lambda_C \sim \Lambda_0$.
If $\Lambda_C \ll \Lambda_0$, we can consider the coupled channel scale
to be light, $\Lambda_C \sim Q$, and the resolvent operator
will be of order $Q$, as expected (a more formal account
can be consulted in Refs.~\cite{Cohen:2004kf,Lensky:2011he}).
On the contrary, if $\Lambda_C \sim \Lambda_0$, then the resolvent operator
for the channel $C \neq \alpha$ will be suppressed by two powers
of $Q/\Lambda_C$, that is, by two orders
in the EFT expansion.

In the particular case of heavy mesons, $\Lambda_C =
\sqrt{2 \mu_{\rm HH} \Delta_Q}$,
where $\mu_{\rm HH}$ is the reduced mass of the system and $\Delta_Q$
is the energy split between different pairs of heavy meson systems.
For the charm sector,
$\Lambda_C \simeq 520\,{\rm MeV}$ for the $\rm DD-DD^*$
and $\rm DD^*-D^*D^*$ pairs and $\Lambda_C \simeq 740\,{\rm MeV}$
for $\rm DD-D^*D^*$.
For the bottom sector, the coupled-channel scales are very similar,
yielding $\Lambda_C \simeq 490\,{\rm MeV}$ and
$\Lambda_C \simeq 700\,{\rm MeV}$ respectively.
The previous estimations indicate that $\Lambda_C$ is similar
to the hard scale of the theory, $\Lambda_0 \sim 0.5-1.0\,{\rm GeV}$,
which indicates that coupled channel effects are suppressed.

A surprising aspect about the coupled channel effects is that 
the associated momentum scale is similar
in the charm and bottom sectors.
This can be easily understood if we consider the heavy quark limit,
$m_Q \to \infty$, in which we expect $\mu_{\rm HH} \sim m_Q$ and
$\Delta_Q \sim 1/m_Q$, meaning that $\Lambda_C \sim m_Q^0$.
That is, the suppression of the coupled channel effects is basically
independent of the heavy quark mass.
Even though in the heavy quark limit the heavy mesons are degenerate,
the energy split between them remains beyond the scope of
a non-relativistic chiral EFT.
The convergence of the EFT description for non-relativistic particles
depends on momentum scales instead of energy scales.
However, from the scaling of the center-of-mass energy with respect to
the heavy quark mass, i.e. $E \sim p^2 / m_Q$,
we see that the energy window in which the EFT is valid decreases
as the heavy quark mass grows.

\section{The Leading Order Potential}
\label{sec:LO}

The lowest order potential in the effective field theory description of
heavy meson molecules can be decomposed into a contact-range and a
finite-range piece
\begin{eqnarray}
V^{(0)}_{H\bar{H}} &=& V^{(0)}_{\rm C} + V^{(0)}_{\rm F} \, , 
\end{eqnarray}
where the subscripts $_C$ and $_F$ are used to denote the contact and
finite-range character of each contribution to the potential.
In the equation above we have labelled the potential by the expected
naive dimensional scaling, rather than the actual scaling
that depends upon the particular power counting
under consideration.
Of course, the existence of a shallow molecular scale will imply that
at least $V_C$ is of order $Q^{-1}$.
However, we will not consider the actual order of the different
contributions of the potential (and their consequences) until
the next section.

\subsection{The Contact Range Potential}

At ${\rm LO}$ we will naively expect a total of six counterterms
in a given isospin channel.
This number correspond to the number of $s$-wave channels
for the ${\rm H\bar{H}}$ system: a $0^{++}$ state for $\rm P\bar{P}$,
two opposite C-parity states, $1^{++}$ and $1^{+-}$
for the $\rm P^*\bar{P}$/$\rm P\bar{P}^*$ system,
and the three $0^{++}$, $1^{+-}$ and $2^{++}$ $\rm P^*\bar{P}^*$ states.
However, the contact range is constrained by the requirements
of heavy quark spin symmetry (HQSS)~\cite{Isgur:1989vq,Isgur:1989ed},
which in turn implies that the six counterterm figure is reduced to only
two independent counterterms~\cite{AlFiky:2005jd}.
We will call these counterterms $C_{0a}$ and $C_{0b}$.
Ignoring the coupled channel effects, the potential reads
(see the Appendix for details)
\begin{eqnarray}
V^{(0)}_{C, {\rm P \bar P}}(\vec{q}, {0^{++}}) &=& C_{0a} \, , 
\end{eqnarray}
\begin{eqnarray}
V^{(0)}_{C, {\rm P^* {\bar P} / P {\bar P}^*}}(\vec{q}, {1^{+-}})
&=& C_{0a} - C_{0b} \, , \\ 
V^{(0)}_{C, {\rm P^* {\bar P} / P {\bar P}^*}}(\vec{q}, {1^{++}})
&=& C_{0a} + C_{0b}\, ,
\end{eqnarray}
\begin{eqnarray}
V^{(0)}_{C, {\rm P^* {\bar P}^* }}(\vec{q}, {0^{++}})
&=& C_{0a} - 2\,C_{0b} \, , \\
V^{(0)}_{C, {\rm P^* {\bar P}^* }}(\vec{q}, {1^{+-}})
&=& C_{0a} - C_{0b} \, , \\
V^{(0)}_{C, {\rm P^* {\bar P}^* }}(\vec{q}, {2^{++}})
&=& C_{0a} + C_{0b} \, , 
\end{eqnarray}
depending on the $J^{PC}$ quantum number and the specific ${\rm H\bar{H}}$ 
system under consideration.
The explicit representation of these two counterterms
in the coupled channel basis of ${\rm H\bar{H}}$ states
can be consulted in Ref.~\cite{Mehen:2011yh}.

\subsection{The Finite Range Potential}

The OPE potential between a heavy meson and anti-meson is local,
and hence it only depends on the exchanged momentum between the heavy mesons:
\begin{eqnarray}
\langle \vec{p}\,' | V^{(0)}_{\rm F} | \vec{p} \rangle =
V^{(0)}_{\rm F}(\vec{q}) \, ,
\end{eqnarray}
with $\vec{q} = \vec{p} - \vec{p}\,'$.
The explicit consideration of the pseudoscalar / vector heavy meson channels
allows us to write the potential as
\begin{eqnarray}
V^{(0)}_{\rm F, PP \to PP}(\vec{q}) &=& 0 \, , \\
V^{(0)}_{\rm F, P^*{P} \to P{P}^* }(\vec{q}) &=& - 
\frac{g^2}{2 f_{\pi}^2}\,\vec{\tau}_1\cdot\vec{\tau}_2\,
\frac{\vec{\epsilon}_1 \cdot \vec{q}\,{\vec{\epsilon}_2\,}^* \cdot \vec{q}}
{q^2 + \mu^2} \, , \\
V^{(0)}_{\rm F,P^*{P}^* \to P^*{P}^*}(\vec{q}) &=& - 
\frac{g^2}{2 f_{\pi}^2}\,\vec{\tau}_1\cdot\vec{\tau}_2\,
\frac{\vec{S}_1 \cdot \vec{q}\,\vec{S}_2 \cdot \vec{q}}{q^2 + m_{\pi}^2} \, ,
\end{eqnarray}
depending on the particular heavy meson channel under consideration.
In the equations above $g$ is the pion axial coupling,
$f_{\pi} = 132\,{\rm MeV}$ the pion decay constant, 
$m_{\pi} = 138\,{\rm MeV}$ the pion mass,
$\vec{\tau}_{1(2)}$ the isospin operators on the heavy meson $1(2)$,
$\vec{\epsilon}_{1(2)}$ is the heavy vector meson polarization,
and $\vec{S}_{1(2)}$ is the spin operator
for intrinsic spin $S = 1$.
For the $\rm P^*{P} \to P{P}^*$ channel, we use $\mu$ instead of $m_{\pi}$:
as a consequence of the different masses of the $P$ and $P^*$ mesons,
the pion is emitted with the zero component of the momentum different
to zero.
In the static limit this is equivalent to changing the effective mass of
the pion.
We have that $\mu^2 = m_{\pi}^2 - \Delta_Q^2$, with $\Delta_Q$ the mass
splitting between the $\rm P$ and $\rm P^*$ mesons.
In the heavy quark limit, $m_Q \to \infty$,
we obtain $\mu = m_{\pi} + \mathcal{O}(1/m_Q)$,
and it would be practical to expand the potential in powers of $1/m_Q$.
While this is the situation we find in the bottom sector ($\rm B\bar{B}^*$),
the charm one (the $\rm D\bar{D}^*$ potential) is more complicated
as we have $\mu^2 < 0$.
The OPE potential thus acquires a small imaginary piece
that is related to the probability loss induced
by the open decay channel $\rm D {\bar D^*} / {\bar D} D^* \to D {\bar D} \pi$.
In this regard it is not clear up to what extend the static approximation
to the potential holds.
The rigorous theoretical treatment of this situation requires the inclusion
of pions as dynamical degrees of freedom and the explicit consideration of
the $\rm D\bar{D} \pi$ three body channel~\cite{Baru:2011rs},
which indicates a significant impact of the three body dynamics
on the $X(3872) \to \rm D\bar{D} \pi$ decay rate, 
but only a mild effect on the $\rm D\bar{D}^*$ wave functions (that is,
the residue of the $X(3872)$ pole in the language of Ref.~\cite{Baru:2011rs}).
This seems to indicate that, if we are only interested in the wave functions,
we can simply ignore the imaginary piece of the OPE potential for $\mu^2 < 0$
and continue using the static limit.

Apart from the previous, there are additional pieces of the potential
that mix the different heavy meson channels.
However, as we have seen, coupled channel effects can be safely disregarded
(even in the heavy quark limit) as their size is similar
to the short range effects beyond chiral symmetry.
Therefore we have decided to ignore the particle coupled-channel terms.

The previous potential has been computed for the meson-meson case.
For obtaining the meson-antimeson potential in the isospin symmetric basis,
we perform a G-parity transformation: this changes the sign of
the $\rm P^*P^*$ potential, but leaves the $\rm P^*P$/$\rm PP^*$ potential
unchanged.
However, the $\rm P^*{\bar P}$/$\rm P{\bar P}^*$ potential is better written
in a definite C-parity basis, for which we employ
\begin{eqnarray}
| P^*\bar{P} (\eta) \rangle = \frac{1}{\sqrt{2}}\,
\left[ | P^*\bar{P} \rangle - \eta\,| P\bar{P}^* \rangle \right] \, ,
\end{eqnarray}
where $\eta$ is the intrinsic C-parity of the meson-antimeson system.
This translates into a factor of $-\eta$
for the $\rm P^*{\bar P}$/$\rm P{\bar P}^*$
potential when expressed in this basis.
The final form of the meson-antimeson potentials is thus
\begin{eqnarray}
V^{(0)}_{\rm F, P^*\bar{P}(\eta)}(\vec{q}) &=& \eta\, 
\frac{g^2}{2 f_{\pi}^2}\,\vec{\tau}_1\cdot\vec{\tau}_2\,
\frac{\vec{\epsilon}_1 \cdot \vec{q}\,{\vec{\epsilon}_2\,}^* \cdot \vec{q}}
{q^2 + \mu^2} \, , \\
V^{(0)}_{\rm F,P^*\bar{P}^*}(\vec{q}) &=&  
\frac{g^2}{2 f_{\pi}^2}\,\vec{\tau}_1\cdot\vec{\tau}_2\,
\frac{\vec{S}_1 \cdot \vec{q}\,\vec{S}_2 \cdot \vec{q}}{q^2 + m_{\pi}^2} \, .
\end{eqnarray}
The general form of the potential can be schematically written as
\begin{eqnarray}
V^{(0)}_{\rm F, {\rm H{\bar H}}}(\vec{q}) = \eta \,
\frac{g^2}{2 f_{\pi}^2}\,\vec{\tau}_1\cdot\vec{\tau}_2\,
\frac{\vec{a}_1 \cdot \vec{q}\,\vec{b}_2 \cdot \vec{q}}{q^2 + \mu^2} \, ,
\end{eqnarray}
where $\vec{a}_1$, $\vec{b}_2$ represent the particular spin operators and
$\mu$ the particular value of the effective pion mass
to be used in each case.
In this form, the intrinsic C-parity $\eta$ is then to be taken $\eta = 1$
for the $P^*\bar{P}^*$ channel.

The coordinate space potential can be obtained by Fourier-transforming
the momentum space one, in which case we obtain
\begin{eqnarray}
\label{eq:OPE-generic}
V^{(0)}_{\rm F, {\rm H\bar{H}}}(\vec{r})
&=& \eta\,\vec{\tau}_1\cdot\vec{\tau}_2\,\vec{a}_1 \cdot \vec{b}_2\,
\frac{g^2}{6 f_{\pi}^2}\,\delta(\vec{r}) 
\nonumber \\ &-&
\eta\,\vec{\tau}_2 \cdot \vec{\tau}_1\,\Big[
{\vec{a}_1} \cdot {\vec{b}_2} W_C (r) \nonumber \\ &+&
(3\,{\vec{a}_1} \cdot \hat{r}\,{\vec{b}_2} \cdot \hat{r} - 
{\vec{a}_1} \cdot {\vec{b}_2}) \, W_T (r) \Big] \, ,
\end{eqnarray}
where $W_C(r)$ and $W_T(r)$ are the central~\footnote{Or, more properly,
the spin-spin part of the potential.}
and tensor pieces of the potential, which read
\begin{eqnarray}
\label{eq:WC}
W_C(r) &=& \frac{g^2 \mu^3}{24 \pi f_{\pi}^2}\,\frac{e^{-\mu r}}{\mu r}\, , \\
\label{eq:WT}
W_T(r) &=& \frac{g^2 \mu^3}{24 \pi f_{\pi}^2}\,\frac{e^{-\mu r}}{\mu r} \, 
\left( 1 + \frac{3}{\mu r} + \frac{3}{\mu^2 r^2} \right) \, .
\end{eqnarray}
The interesting feature of the tensor force is
that it can mix channels with different angular momentum.

\subsection{The Partial Wave Decomposition}

In this subsection we will show the explicit partial wave decomposition
of the potential.
We will use the spectroscopic notation $^{2S+1}L_J$ for characterizing
a certain partial wave with spin $S$, orbital angular momentum $L$
and total angular momentum $J$.
If two or more partial waves are coupled we will indicate it with a dash,
for example $^3S_1$--${}^3D_1$, $^5P_2$--${}^5F_2$ or
$^1F_3$--${}^5P_3$--${}^5F_3$--${}^5H_3$.

There are certain rules on how partial waves coupled to each other.
On the one hand, the central operator, which we define as
\begin{eqnarray}
C^{{\rm P\bar{P}^*}}_{12}(\hat{r}) &=& 
{\vec{\epsilon}_1} \cdot {\vec{\epsilon}_2^{\,\,*}} \, , \\
C^{{\rm P^*\bar{P}^*}}_{12}(\hat{r}) &=& {\vec{S}_1} \cdot {\vec{S}_2} \, ,
\end{eqnarray}
conserves parity, C-parity, spin, orbital and total angular momentum.
Consequently it does not mix partial waves.
On the other hand, the tensor operator, given by
\begin{eqnarray}
S^{{\rm P\bar{P}^*}}_{12}(\hat{r}) &=& 
3\,{\vec{\epsilon}_1} \cdot \hat{r}\,{\vec{\epsilon}_2^{\,\,*}} \cdot \hat{r} - 
{\vec{\epsilon}_1} \cdot {\vec{\epsilon}_{2}^{\,\,*}} \, , \\
S^{{\rm P^*\bar{P}^*}}_{12}(\hat{r}) &=& 
3\,{\vec{S}_1} \cdot \hat{r}\,{\vec{S}_2} \cdot \hat{r} - 
{\vec{S}_1} \cdot {\vec{S}_2} \, ,
\end{eqnarray}
conserves parity, C-parity and total angular momentum,
but neither spin nor orbital angular momentum.
However, parity and C-parity conservation imply that 
the tensor force can only change
the spin $S$ and orbital angular momentum $L$ by an even number of units.
This in turn implies that we have four possibilities,
which we define as follows
\begin{eqnarray}
{}^3U_J &\equiv& {}^3J_J \, , \\
{}^3C_J &\equiv& {}^3(J-1)_J - {}^3(J+1)_J \, , \\
{}^5C_J &\equiv& {}^5(J-1)_J - {}^5(J+1)_J \, , \\
{}^{1-5}C_J &\equiv& {}^1J_J - {}^5(J-2)_J - {}^5J_J - {}^5(J+2)_J \, ,
\end{eqnarray}
where ($U$)$C$ stands for (un)coupled.
In this notation, $^3C_1$ is $^3S_1$--${}^3D_1$, $^5C_2$ stands for
$^5P_2$--${}^5F_2$ and $^{1-5}C_3$ would be
$^1F_3$--${}^5P_3$--${}^5F_3$--${}^5H_3$.
Of course, low $J$ coupled channels can be uncoupled or have less members
than expected.
Examples are the $^3P_0$ ($^3C_0$) or the $^1S_0$--${}^5D_0$ ($^{1-5}C_0$)
partial waves.

The matrix elements of the central and tensor operators between different
$^{2S+1}L_J$ partial wave are schematically calculated as
\begin{eqnarray}
\langle b | \hat{O}_{12} | a  \rangle &=&
\delta_{J_b J_a}\delta_{M_b M_a}\,
\sum_{\{\mu\}} \int d^2\hat{r}\, 
\langle {\mu_b} | \hat{O}_{12}(\hat{r}) | {\mu_a} \rangle 
\nonumber \\
&=& \delta_{J_b J_a}\delta_{M_b M_a}\, O_{J (S_b S_a L_b L_a)} \, ,
\end{eqnarray}
where $a$, $b$ are the initial and final partial wave, characterized
by the vector $| (S_{a(b)} L_{a(b)}) J_{a(b)} M_{a(b)} \rangle$,
and $\{ \mu \}$ is whatever internal angular momentum
quantum numbers $a$ and $b$ have.
The two Kronecker $\delta$'s are a consequence of
total angular momentum conservation.
In the case of the central operator, which also conserves $S$ and $L$,
we simply have
\begin{eqnarray}
C_{J (S_b S_a L_b L_a)} &=& C_{J(SL)} \delta_{S_b S_a}\,\delta_{L_b L_a} \, , 
\end{eqnarray}
with
\begin{eqnarray}
C^{P\bar{P}^*}_{J(1 L)} &=& 1 \, , \\
C^{P^*\bar{P}^*}_{J(S L)} &=& \frac{1}{2}\,\left[ S(S+1) - 4 \right] \, , 
\end{eqnarray} 
in the $\rm P\bar{P}^*$ and $\rm P^*\bar{P}^*$ cases respectively.

We write the matrix elements of the tensor operator directly
in one of the four coupled channel basis previously defined. 
For the uncoupled channel case we have
\begin{eqnarray}
\label{eq:S12_PPst_3UJ}
{S}^{P\bar{P}^*}_{J}({}^3U_J) &=& - 1 \, , \\
{S}^{P^*\bar{P}^*}_{J}({}^3U_J) &=& + 1\, .
\end{eqnarray}
In the $^3C_J$ case we obtain
\begin{eqnarray}
{S}^{P\bar{P}^*}_{J=0}({}^3C_J) &=& -2 \, , \\
{S}^{P^*\bar{P}^*}_{J=0}({}^3C_J) &=& +2 \, ,
\end{eqnarray}
\begin{eqnarray}
{\bf S}^{P\bar{P}^*}_{J \geq 1}(^3C_J) &=& + \frac{1}{2 J + 1} \nonumber \\
&\times& \begin{pmatrix}
J - 1  &  -3 \sqrt{J(J+1)} \\
-3 \sqrt{J(J+1)} & J + 2
\end{pmatrix} \, , \nonumber \\
\\
{\bf S}^{P^* \bar{P}^*}_{J \geq 1}(^3C_J) &=& - \frac{1}{2 J + 1} \nonumber \\
&\times& \begin{pmatrix}
J - 1  &  -3 \sqrt{J(J+1)} \\
-3 \sqrt{J(J+1)} & J + 2
\end{pmatrix} \, , \nonumber \\
\end{eqnarray}
The other two set of coupled channels, $^5C_J$ and $^{1-5}C_J$,
only happen for the $\rm P^* \bar{P}^*$ system.
In the $^5C_J$ case, we obtain the matrices
\begin{eqnarray}
{S}^{P^* \bar{P}^*}_{J=1}(^5C) &=& -1 \, , \\
{\bf S}^{P^* \bar{P}^*}_{J \geq 2}(^5C) &=& \frac{1}{2J+1} \nonumber \\
&\times&
\begin{pmatrix}
J+5 & 3\sqrt{J(J+1) - 2} \\
3\sqrt{J(J+1) - 2} & J-4
\end{pmatrix} \, , \nonumber \\
\end{eqnarray}
where there is no $J = 0$ partial wave with fits into
the $^5C_J$ scheme (the $^5P_0$ wave is unphysical).
In the $^{1-5}C_J$ case, we obtain
\begin{eqnarray}
{\bf S}^{P^* \bar{P}^*}_{J=0}(^{1-5}C) &=&
\begin{pmatrix}
0 & -\sqrt{2} \\
-\sqrt{2} & -2
\end{pmatrix} \, , \\
{\bf S}^{P^* \bar{P}^*}_{J=1}(^{1-5}C) &=& 
\begin{pmatrix}
0 & \frac{2}{\sqrt{5}} & -\sqrt{\frac{6}{5}} \\
\frac{2}{\sqrt{5}} & -\frac{7}{5} & \frac{\sqrt{6}}{5}\\
-\sqrt{\frac{6}{5}} & \frac{\sqrt{6}}{5} & -\frac{8}{5}
\end{pmatrix} \, , 
\end{eqnarray}
\begin{widetext}
\begin{eqnarray}
\label{eq:S12_PstPst_15CJ}
{\bf S}^{P^* \bar{P}^*}_{J \geq 2}(^{1-5}C) &=& 
\begin{pmatrix}
0 & -\sqrt{\frac{3 J (J-1)}{(2J+1)(2J-1)}} & \sqrt{\frac{2 J (J+1)}{(2J+3)(2J-1)}} & -\sqrt{\frac{3 (J+1) (J+2)}{(2J+3)(2J+1)}} \\
-\sqrt{\frac{3 J (J-1)}{(2J+1)(2J-1)}} & -\frac{2J-4}{2J-1} &
\sqrt{\frac{6 (J+1) (J-1) (2J + 3)}{(2J+1){(2J-1)}^2}} & 0\\
\sqrt{\frac{2 J (J+1)}{(2J+3)(2J-1)}} &
\sqrt{\frac{6 (J-1) (J+1) (2J+3)}{(2J+1){(2J-1)}^2}} 
& \frac{(2J+5)(2J-3)}{(2J+3)(2J-1)} &
\sqrt{\frac{6 J (J+2) (2J-1)}{{(2J+3)}^2 (2J +1)}}\\
-\sqrt{\frac{3 (J+1) (J+2)}{(2J+3) (2J+1)}} & 0 & 
\sqrt{\frac{6 J (J+2) (2J -1)}{{(2J+3)}^2 (2J+1)}} & 
-\frac{2J+6}{2J+3}
\end{pmatrix} \, .
\end{eqnarray}
\end{widetext}
For $J=0, 1$ the number of angular momentum channels is smaller than expected
and the corresponding tensor matrices are written in the $^1S_0$--${}^5D_0$ 
basis for $J = 0$ and $^1P_1$--${}^5P_1$--${}^5F_1$ for $J = 1$.

\section{The Power Counting Map for Heavy Meson Molecules}
\label{sec:pc}

The presence of a bound state between a heavy meson and antimeson
indicates the necessity of promoting a piece of the EFT potential from
order $Q^0$ to $Q^{-1}$.
If the molecular state is shallow,
as happens in the $X(3872)$ or the $Z_b(10610)$ and $Z_b(10650)$,
probably it is enough to promote a contact range interaction only.
However, if the state is sufficiently deep (exactly how deep will be
the subject of discussion of the next section), we are required to
additionally promote the OPE potential as well.
In this second case, the renormalizability of the theory will in turn
generate important changes in the counterterm structure already
at ${\rm LO}$, as has been repeatedly discussed
in the EFT description of nuclear forces~\cite{Nogga:2005hy,Birse:2005um,Valderrama:2009ei,Valderrama:2011mv,Long:2011qx,Long:2011xw}.
Therefore we distinguish two cases:
power counting with perturbative and non-perturbative pions.

\subsection{Counting with Perturbative Pions}

We begin by considering a theory in which there is at least a $Q^{-1}$
contact operator but where pions are perturbative.
In this case, pions represent a ${\rm NLO}$ correction.
As there are two contact operators at ${\rm LO}$, $C_{0a}$ and $C_{0b}$,
we have three options in what pertains to operator promotions,
defining three power counting schemes:
\begin{itemize}
\item[{(a)}] $C_{0a}$ is of order $Q^{-1}$,
\item[{(b)}] $C_{0b}$ is of order $Q^{-1}$,
\item[{(c)}] both $C_{0a}$ and $C_{0b}$ are of order $Q^{-1}$.
\end{itemize}
Power countings (a), (b) and (c) are the EFT restatement of
the observation made by Voloshin~\cite{Voloshin:2011qa}
about the heavy quark spin structure of the $Z_b$ resonances:
depending on which piece of the heavy quark spin symmetric interaction
is responsible of the appearance of the two low-lying $Z_b$ states,
we should expect a total of four or six
molecular s-wave states of $\rm B^{(*)}\bar{B}^{(*)}$.
The later situation, six states, correspond to the promotion (a),
while the former, giving four states, with (b)~\footnote{Strictly speaking
in (b) we should only expect three low-lying states,
the two $1^{+-}$ states and the $\rm P^*\bar{P}^*$ $0^{++}$ state.
However, if we recover the coupled channel effects,
a fourth $\rm P\bar{P}$ $0^{++}$ state may be expected.
In this regard, it is more natural to use the promotion (c)
to obtain the four states.
} and (c).
It is worth commenting that the third possibility (c) is quite general
and can also accommodate other situations.
For example if $C_{0a} < 0$, $C_{0b} > 0$ and $C_{0a} + C_{0b}$
is negative and attractive enough as to bind, then there will
be six states, but at most only two of them will
be shallow (the $1^{++}$ and $2^{++}$).
However, this specific situation does not necessarily correspond
to what we encounter in the $Z_b(10610)$ and $Z_b(10650)$ resonances.

Power countings (a) and (b/c) also differ with respect to the treatment
of coupled channel effects.
The contact range interaction responsible for mixing different particle
channels is $C_{0b}$.
In contrast, $C_{0a}$ is diagonal in the ${\rm H \bar H}$ space.
This means that in the (b/c) counting the coupled channels effects
are promoted by
one order in the EFT expansion, and are therefore ${\mathcal O}(Q^1)$,
instead of the naive estimate ${\mathcal O}(Q^2)$.
In turn, the coupled channel effects dependent on the OPE interaction
will still be of order $Q^2$.

In what regards the subleading contact operators, $C_{2n}$,
their scaling is affected by the corresponding scaling of $C_{0}$.
By this we mean that if $C_0$ is $Q^{-1}$ ($Q^0$),
then $C_{2n}$ is $Q^{2n-2}$ ($Q^{2n}$),
as can be deduced from renormalization group analysis
arguments~\cite{Birse:1998dk}.
Of course, only the $C_{2n}$ operators with a similar HQSS structure 
as their $Q^{-1}$ counterparts will be promoted,
depending on the whether we are considering
the (a), (b) or (c) countings.
However, the point is that, independently of the kind of promotion,
we will have new free parameters at ${\rm NLO}$.
This also implies that without new data to fix the $C_{2}$'s
we cannot proceed to ${\rm NLO}$.
Finally, the expected relative accuracy of a ${\rm LO}$ calculation
is of order $Q/\Lambda_0$.

There is still one contribution that we have not addressed: the effect of
heavy quarkonia states with the adequate quantum numbers in the vicinity 
of any of the ${\rm H \bar H}$ thresholds.
This situation may happen in the $X(3872)$ if the $\chi_{c1}(2S)$ lies nearby,
as has been discussed in Ref.~\cite{Braaten:2010mg}.
However, the coupling to heavy quarkonia is surely going to represent
a subleading contribution only.
The natural expectation is that the counterterm mixing these two states
is of order $Q^0$.
Therefore the coupled channel effects will be of order $Q^0$ and
hence subleading if the thresholds of the $X(3872)$ and
the $\chi_{c1}(2S)$ are very close,
or order $Q^2$ otherwise.
This type of contribution, being subleading, does not really
affect the ${\rm LO}$ calculation and its accuracy.

\subsection{Counting with Non-Perturbative Pions}

As we will see in the next section, pion exchanges can eventually become
non-perturbative above a certain critical energy
in heavy meson-antimeson systems.
Interestingly, the non-perturbative treatment of pion exchange
will be able to significantly alter the power counting scheme,
as happens in the two-nucleon system~\cite{Nogga:2005hy,Birse:2005um,Valderrama:2009ei,Valderrama:2011mv,Long:2011qx,Long:2011xw}.
We notice that Ref.~\cite{Nieves:2011zz} already develops
an EFT description of heavy meson molecules
with non-perturbative OPE. 
However, the previous work only considers the particular case of
the ${\rm P^* \bar P}$/${\rm P \bar P^*}$ channels.
In the next lines we will concentrate instead on the general HQSS
structures relating the different $\rm H \bar H$ particle channels.

We must begin with distinguishing which piece of the OPE potential
is to be iterated, the central or the tensor.
If only the central pieces is non-perturbative,
the changes in the power counting are minimal
with respect to the theories in which we already
iterate the contact range operators,
as demonstrated by Barford and Birse~\cite{Barford:2002je}.
The reason is that the $1/r$ singularity of central OPE is not enough
to change the power-law behaviour of the wave function
at short distances, which is intimately related
to the scaling of the counterterms
(see, for example, the detailed discussion of Ref.~\cite{Valderrama:2011mv}).
In such a case, we simply refer to the results of the previous section.

On the contrary, the non-perturbative treatment of tensor OPE entails
significant changes in the scaling of the contact range operators,
as has been extensively discussed in the context
of nuclear EFT~\cite{Nogga:2005hy,Birse:2005um,PavonValderrama:2005uj,PavonValderrama:2005wv,Valderrama:2009ei,Valderrama:2011mv,Long:2011qx,Long:2011xw}.
The reason is that tensor OPE is a singular potential
behaving as $1/r^3$ at distances below the pion Compton wavelength.
The most evident effect of a singular potential is on the ${\rm LO}$
counterterms: if the singular potential is attractive,
the renormalizability of the theory requires
the inclusion of a contact interaction
to stabilize the cut-off dependence~\cite{PavonValderrama:2005uj,PavonValderrama:2005wv}.
This can be appreciated if we consider the solution of
the reduced Schr\"odinger equation for a $-1/r^3$
potential, which for short enough distances
read~\cite{PavonValderrama:2005uj,PavonValderrama:2005wv}:
\begin{eqnarray}
u_k(r) \to {\left( \frac{r}{a_3} \right)}^{3/4}\,\sin{\left[
\frac{a_3}{r} + \varphi \right]} \, .
\end{eqnarray}
The problem with the wave function above is that it is always regular
at the origin, no matter what the values of the semiclassical phase
$\varphi$ is.
If we try to regularize the potential by cutting it off at short distances,
i.e. $V(r; r_c) = V(r)\,\theta (r - r_c)$, we do obtain a value
of $\varphi (r_c)$.
However, there is no well-defined $r_c \to 0$ limit for $\varphi (r_c)$:
the semiclassical phase simply oscillates faster and faster on its way
to the origin.
The solution is to make the counterterm oscillate~\cite{PavonValderrama:2010fb},
\begin{eqnarray}
\frac{\mu_R C_0(r_c)}{2 \pi r_c^2} &\to& -\frac{2}{a_3}\,
{\left( \frac{r}{a_3} \right)}^{3/4}\,
\cot{\left[ \frac{a_3}{r} + \varphi \right]} \, ,
\end{eqnarray}
so $\varphi$ remains a constant.
In contrast, if the potential is repulsive, we have a unique and regular
solution at short distances
\begin{eqnarray}
u_k(r) \to {\left( \frac{r}{a_3} \right)}^{3/4}\,\exp{\left[ -
\frac{a_3}{r} \right]} \, ,
\end{eqnarray}
and no counterterm is required~\cite{PavonValderrama:2005uj,PavonValderrama:2005wv}.

The (angular momentum) coupled channel case can  be
a bit more complicated however.
In this case the important factor are the eigenvalues of the potential
matrix in coupled channel space:
if there are $n$ attractive (i.e. negative) eigenvalues,
we will need $n(n+1)/2$ counterterms~\cite{PavonValderrama:2005uj,PavonValderrama:2005wv}.
For the tensor matrices we have calculated in the previous section,
the eigenvalues are
\begin{eqnarray}
{\lambda}_{\rm P\bar{P}}(0^{++}) &=& 0 \, ,
\end{eqnarray}
\begin{eqnarray}
{\lambda}_{\rm P\bar{P}^*}(1^{++}) &=& \tau\, \{ 1, -2 \} \, , \\
{\lambda}_{\rm P\bar{P}^*}(1^{+-}) &=& \tau\, \{ 2, -1 \} \, , 
\end{eqnarray}
\begin{eqnarray}
{\lambda}_{\rm P^*\bar{P}^*}(0^{++}) &=& \tau\, 
\{ 1 +\sqrt{3}, 1 -\sqrt{3} \} \, , \\
{\lambda}_{\rm P^*\bar{P}^*}(1^{+-}) &=& \tau\, \{ 2, -1 \} \, , \\
{\lambda}_{\rm P^*\bar{P}^*}(2^{++}) &=& \tau\, 
\{ 1, -2, 1+\sqrt{3}, 1-\sqrt{3} \} \, , 
\end{eqnarray}
with $\tau = \vec{\tau}_1 \cdot \vec{\tau}_2$, that is,
we have included the $- \eta \, \vec{\tau}_1 \cdot \vec{\tau}_2$ factor
multiplying the potential of Eq.~(\ref{eq:OPE-generic})
for obtaining the overall sign right.
From the eigenvalues above it is apparent that
the $C_{0a}$ and $C_{0b}$ counterterms are not enough
as to renormalize the scattering amplitude if tensor
OPE is non-perturbative.

Each distinct negative eigenvalue requires a different,
independent counterterm.
However, if two different channels share a negative eigenvalue,
a common counterterm will be able to renormalize
the two channels~\cite{PavonValderrama:2010fb}.
In principle we can count a total of eight counterterms for renormalizing
the six s-wave states (one per channel, with the exception of the $2^{++}$
that requires three counterterms).
This figure is then reduced by counting the shared eigenvalues, that is,
(i) the $1^{+-}$ ${\rm P^* \bar P}$ and ${\rm P^* \bar P^*}$ molecules
can be renormalized with the same counterterm, (ii) the $0^{++}$
and $2^{++}$ ${\rm P^* \bar P^*}$ share another eigenvalue and
finally (iii) another common contact operator between
the $1^{++}$ and $2^{++}$ channels.
In total we end up with five independent counterterms.

All this makes the theory with non-perturbative pions rather cumbersome.
In addition, with five contact interactions HQSS, although still there,
is not so manifest as in the pionless theory~\footnote{What is actually
happening is that certain s- to d-wave operators are being promoted
from $Q^2$ to $Q^{-1}$. However this may require the additional
promotion of all (s- to s-wave) counterterms
with two derivatives. Moreover, non-perturbative pions can change
the bound state spectrum considerably from the expectations
based on the HQSS of the potential,
see Ref.~\cite{Ohkoda:2011vj}
for an explicit example. 
A further issue, which we do not discuss in this work,
is the possibility to go beyond the additivity of
the long and short range forces, see footnote 6 of Ref.~\cite{Entem:2007jg}.
}.
It is interesting to notice though that the $1^{+-}$ ${\rm P^* \bar P}$
and ${\rm P^* \bar P^*}$ states are still expected to be degenerate:
this situation corresponds to the quantum numbers of the $Z_b(10610)$
and $Z_b(10650)$ resonances.
Luckily, as we will see in the next section, non-perturbative tensor forces
are only expected in the isoscalar $\rm B^{(*)}\bar{B}^{(*)}$ states.

However the non-perturbative OPE counting has certain advantages too.
The momentum dependent $C_2$ operators are not expected until
order $Q^{3/2}$ at least (or even $Q^2$: this is still
an unresolved issue in nuclear EFT~\cite{Valderrama:2011mv}).
The lowest subleading order corrections are expected to happen
at order $Q^{1}$:
they are subleading contributions to the $C_0$ operators
that are needed for absorbing the expected divergences
in (particle) coupled channels,
owing to the $1/r^3$ singularity of the tensor force.
Thus the expected relative error of a ${\rm LO}$ calculation
is now $(Q/\Lambda_0)^2$, representing an improvement over
the $(Q/\Lambda_0)$ error in the perturbative pion case.

\section{The Perturbative Treatment of One Pion Exchange}
\label{sec:pert}

\subsection{Central One Pion Exchange}

\begin{figure}[htb]
\begin{center}
\includegraphics[height=3.5cm]{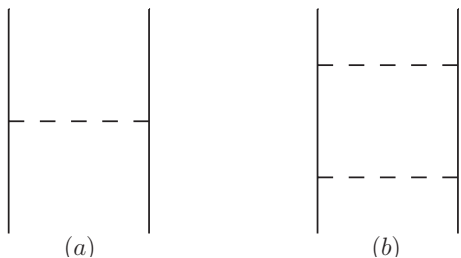}
\end{center}
\caption{
Feynman diagrams corresponding to (a) the OPE potential between two heavy
mesons and (b) the first iteration of the OPE potential.
The ratio of diagram (b) over (a) can be used to determine
the energy range in which OPE is perturbative.
}
\label{ope-ratio}
\end{figure}

The perturbative character of the central piece of the OPE potential
can be determined by a direct comparison of the diagrams
of Fig.~(\ref{ope-ratio}).
This corresponds to computing the ratio of the matrix elements of
the operators $V^{(0)} G_0 V^{(0)}$ and $V^{(0)}$, 
which in terms of power counting is expected to scale as
\begin{eqnarray}
\frac{\langle p | V^{(0)} G_0 V^{(0)} | p \rangle}
{\langle p | V^{(0)} | p \rangle} \sim \frac{Q}{\Lambda_{\rm C}} \, ,
\end{eqnarray}
indicating the breakdown scale of a theory with central perturbative pions,
where it is understood that we only consider the central piece of $V^{(0)}$.
The evaluation is trivial for $s$-waves and $p = 0$,
for which the generic low energy scale $Q$ can only be identified
with the pion mass $m_{\pi}$, leading to~\cite{Fleming:1999ee}
\begin{eqnarray}
\Lambda_{\rm C} &=& \frac{1}{| \sigma\,\tau |}
\frac{24 \pi f_{\pi}^2}{\mu_{\rm H \bar H} \, g^2} \, 
\end{eqnarray}
where $\sigma = \frac{1}{2}\,S(S+1) - 2$ and $\tau = 2\,I(I+1) - 3$ 
with $S$ and $I$ the total spin and isospin of the system.
In general, this number is quite large for isovector channels
($\Lambda_C \gtrsim 1\,{\rm GeV}$),
so the central piece of the OPE potential can always be
treated as a perturbation in these cases.
In contrast, for the $S=0$ isoscalar case, corresponding to a
$I^G(J^{PC}) = 0^+(0^{++})$ $P^*\bar{P}^*$ molecule,
the central piece can be quite important,
but not so much as the tensor piece,
as we will see later.

\subsection{Tensor One Pion Exchange}

The determination of the breakdown scale for a power counting
in which tensor OPE is order $Q^0$ and hence perturbative
is not trivial.
The reason is that the argument of comparing the relative size of diagrams
does not work when considering the tensor piece of OPE.
In first place, tensor OPE does not directly act on $s$-waves,
but only indirectly owing to transitions to
intermediate $d$-wave states.
This means that we should consider second and third order
perturbation theory to obtain the breakdown scale
\begin{eqnarray}
\label{eq:ratio-tensor}
\frac{\langle V_{sd}^{(0)} G_0 V_{dd}^{(0)} G_0 V_{ds}^{(0)} \rangle}
{\langle V_{sd}^{(0)} G_0 V_{ds}^{(0)} \rangle} \sim \frac{Q}{\Lambda_{\rm T}}
\, ,
\end{eqnarray}
where by $V^{(0)}$ we refer to the tensor piece only.
However, there is a serious problem at this point:
in the equation above the numerator and denominator are divergent.
The reason lies in the $1/r^3$ divergent behaviour of the tensor force
at distances below the Compton wavelength of the pion, which induces
a linear (quadratic) divergence in the denominator (numerator),
see for example Ref.~\cite{PavonValderrama:2005gu}
for the details of this kind of calculations.
From the power counting point of view, $V^{(0)} G_0 V^{(0)}$ requires
the inclusion of the order $Q$ correction to the $C_0$ counterterm,
while $V^{(0)} G_0 V^{(0)} G_0 V^{(0)}$ involves in addition
two new order $Q^2$ counterterms, a purely $s$-wave one
and another one connecting the $s$- and $d$-wave channels
for the result to be properly renormalized~\footnote{
Of course, this is assuming that the leading counterterm appears
at order $Q^0$. If we are considering instead a two heavy meson
system with a shallow bound state, then the leading counterterm
is order $Q^{-1}$ and the $C_2$/$C_4$ of order $Q^0$/$Q^2$.}.
The divergence spoils the evaluation of the previous ratio.
If we include counterterms, the ratio is again finite, but is also
contaminated by whatever we assume to be the hard scale
of the counterterms.
In principle it is still possible to find the value of $\Lambda_{T}$
{\it a posteriori} by means of a full $Q^2$ calculation
as in the nuclear case~\cite{Fleming:1999ee},
but that requires plenty of experimental information about the low energy
scattering of heavy mesons to fix
the counterterms.
Therefore we need to resort to other kind of arguments.

The failure of standard perturbation theory for inverse power-law
potentials ($1/r^n$) is well-known in atomic physics,
where techniques have been developed to handle
this type of potentials.
Of particular interest in this regard is the work of
Cavagnero~\cite{Cavagnero:1994zz},
who pointed out that the divergences of the perturbative series
are analogous to the role of {\it secular} perturbations
in classical mechanics, that is, small perturbations that
however end up diverging at large enough time scales.
The solution is to formulate a secular series
in which we redefine (or renormalize) some quantity,
in this case the angular momentum,
in order to obtain finite results at arbitrary order.
By this we specifically mean that the zeroth order perturbative approximation
to the $l$-wave radial wave function is taken to be
\begin{eqnarray}
\Psi_l(r; k) &=& \frac{J_{\nu(k a)}(k r)}{\sqrt{r}} + \mathcal{O}(g) \, ,
\end{eqnarray}
instead of the customary solution $r^{-1/2}\,J_{l+\frac{1}{2}}(k r)$,
where $\nu(k a)$ is the renormalized angular momentum,
$k$ the center-of-mass momentum,
$a$ is the coupling constant and
$J_n(x)$ the Bessel function of order $n$.
The interesting feature is that at low energies the secular expansion
can be reinterpreted as a particular resummation of
the perturbative expansion, i.e.
the renormalized angular momentum $\nu(k a)$ can be expanded
as a power series in $a$.
This observation, if translated to the EFT potential between heavy mesons,
indicates the existence of an energy range for which
the inverse power-law potential can be treated
perturbatively at long distances~\footnote{In this statement
we are implicitly assuming the presence of exponential suppression
($e^{-m r}$) for standard perturbation theory to make sense
at long distances and of suitable counterterms to absorb
the divergences we commented previously.}.

For the particular inverse power-law potential of interest in this work,
the $1/r^3$ potential (corresponding to the form of
the tensor force for $m_{\pi } r < 1$),
the secular series for the wave function have been studied and analyzed
in detail by Gao~\cite{Gao:1999zz} for the uncoupled channel case.
These techniques were extended by Birse~\cite{Birse:2005um}
to the coupled channel case, with a particular emphasis
on the applications to nucleon-nucleon scattering.
In the present work, we particularize the results of Birse~\cite{Birse:2005um}
to the specific type of coupled channels appearing in two heavy meson systems.

In agreement with EFT expectations, the secular expansion
of the wave function converges fast
at long distances / low energies.
Meanwhile, at high enough energies a very interesting thing happens:
the secular expansion becomes a non-converging series in powers of
the coupling constant above a certain critical value of
the center-of-mass momentum~\footnote{Specifically
$\nu(x)$ encounters a non-analyticity in $x = k\,a$.}.
That is, the secular expansion can no longer be considered
as a perturbative series in disguise.
From the EFT viewpoint this critical momentum is to be identified
with the hard scale of the theory:
if we map the secular series of the wave function at low energies
onto the EFT expansion
\begin{eqnarray}
\Psi_{\rm H \bar H} = \sum_{\nu = -1}^{\infty}
\hat{\Psi}_{\rm H \bar H}^{(\nu)}\,{\left( \frac{Q}{\Lambda_T} \right)}^{\nu} \, ,
\end{eqnarray}
it is apparent that the mapping will stop being valid at $Q = \Lambda_T$,
where $Q$ is to be understood as the momentum (as the pion mass is fixed).
This means that $\Lambda_T$ is the tensor breakdown scale
we were looking for.
Above this critical value of the momentum the EFT description of
the two heavy meson system will require the full
iteration of the $1/r^3$ tensor force.

In the following we will consider the OPE potential in the $\mu \to 0$ limit,
for which it simplifies to a pure power-law potential of the type $1/r^3$:
\begin{eqnarray}
2\,\mu_{\rm H\bar{H}}\,{\bf V}^{(0)}_{\rm F}(r) =
\frac{a_3}{r^3}\,{\bf S}_j \, ,
\end{eqnarray}
where ${\bf S}_j$ is the tensor force matrix, $a_3$ is a length scale related
to the strength of the tensor force, and the subscript
$_{\rm H\bar{H}}$ denotes any combination of heavy meson and
antimeson for which the OPE potential is non-zero.
As can be appreciated, only the tensor piece of the OPE potential
survives in this limit.
Even though $\mu = 0$ can be physically interpreted as taking the chiral
and heavy quark limit simultaneously ($m_{\pi} \to 0$ and $m_Q \to \infty$),
here we simply consider the $\mu \to 0$ limit as a mathematical convenience
that allows us to directly apply the results of
Refs.~\cite{Gao:1999zz,Birse:2005um}.
Deviations from $\mu = 0$ will in general rise the value of the critical
momentum, as they will reduce the strength of the potential
at distances $\mu r \geq 1$ and therefore increase the
perturbative character of the interaction
at long distances.
Analogously, the finite size of the heavy mesons is expected to weaken
the potential at short distances.
In this regard, the $\mu = 0$ (plus $\Lambda_0 = \infty$) determination
are merely a lower bound: the real critical momenta will be higher.
We will estimate the finite pion mass corrections to the critical momenta
at the end of this section.

Curiously there is a case for which the condition $\mu = 0$ is almost
fulfilled: the $X(3872)$, in which $m_{\pi}$ and $\Delta_Q$ are very similar.
In such a case the critical momenta calculated in this work
will really represent the actual boundary to
the perturbative treatment of tensor OPE.

\subsubsection{The Uncoupled Channel Case}

We start with the uncoupled channel case, for which the tensor operator
${\bf S}_j = S_j$ is simply a number
(instead of a matrix in the coupled channel basis).
In this case,  the reduced Schr\"odinger equation reads
\begin{eqnarray}
\label{eq:schroe_r3_u}
- u_{k,j}'' + \left[ S_j\,\frac{a_3}{r^3}
+ \frac{l_j (l_j+1)}{r^2} \right]\,u_{k,j}(r) = k^2 \, u_{k,j}(r) \, ,
\nonumber \\
\end{eqnarray}
where $u_{k,j}$ is the reduced wave function,
$l_j$ is the angular momentum of the two body system,
which for the uncoupled channel case coincides with the total
angular momentum of the system ($l_j = j$),
and $k$ is the center-of-mass momentum.
We have $| S_j | = 1$ in all the ${\rm H\bar{H}}$ uncoupled channels,
with the exception of the $^3P_0$ partial wave
for which $| S_j | = 2$.

The general solution of this equation (for $k \neq 0$~\footnote{For simplicity
we are expressing $u_{k,j}$ in terms of wave functions which do not have
a well-defined $k \to 0$ limit.
This limitation can be easily overcome by adding energy-dependent
normalization coefficients, see Ref.~\cite{Gao:1999zz}
for the details of such a normalization.
})
can be written as the linear combination 
\begin{eqnarray}
\label{eq:uk_exp}
u_{k,j}(r) = \alpha\,\xi_{j}(r; k) + \beta\,\eta_{j}(r; k) \, ,
\end{eqnarray}
where $\xi_{k,j}(r)$ and $\eta_{k,j}(r)$ are 
expressible as series in terms of Bessel functions
\begin{eqnarray}
\label{eq:xi_exp}
\xi_{j}(r; k) &=& \sum_{m = -\infty}^{+\infty}\,b_m\,\sqrt{r}\,
J_{m + \nu}(k r) \, , \\
\label{eq:eta_exp}
\eta_{j}(r; k) &=& \sum_{m = -\infty}^{+\infty}\,(-1)^m\,b_{m}\,\sqrt{r}\,
J_{-m - \nu}(k r) \, .
\end{eqnarray}
The shift $\nu$ is a function of the dimensionless parameter $\kappa = k\,a_3$.
The $b_m$ coefficients can be expressed in terms of a recurrence relation
that admits an explicit solution~\cite{Gao:1999zz}.

The dependence of the renormalized angular momentum $\nu$
in terms of $\kappa$ is determined
by the condition that $\nu$ is the zero of a characteristic function
\begin{eqnarray}
F_l(\nu, \kappa) = 0 \, ,
\end{eqnarray}
where the function $F_l(\nu, \kappa)$ is in turn  defined by
\begin{eqnarray}
F_l(\nu, \kappa) \equiv (\nu^2 - \nu_0^2) - \frac{\kappa^2}{\nu}\,
\left[ R_0(\nu) - {R}_0(-\nu)\right] \, ,
\end{eqnarray}
where $\nu_0 = (l_j + \frac{1}{2})$.
In the equation above,
$R_0(\nu)$ can be expressed again in terms of a recursive relation
\begin{eqnarray}
R_n(\nu) = \frac{1}{(\nu + 1)\,[(\nu + 1)^2 - \nu_0^2] - 
S_j^2\,\kappa^2\,R_{n+1}(\nu)} \, . \nonumber \\
\end{eqnarray}
Birse~\cite{Birse:2005um} notices that between 20 and 30 iterations
are enough as to obtain $\nu(\kappa)$ with six significant digits.
As already commented, 
the analytical properties of the function $\nu(\kappa)$ are
the essential ingredient for determining the range of applicability
of the perturbative reexpansion
of the wave function defined in Eqs.~(\ref{eq:uk_exp}),
(\ref{eq:xi_exp}) and (\ref{eq:eta_exp}).

The behaviour of the solutions $\nu = \nu(\kappa)$ can be described as follows:
for $\kappa = 0$ (plus the condition $k \neq 0$)
we have the Schr\"odinger equation for free waves with angular momentum $l$.
In such a case, the solution of Eq.~(\ref{eq:schroe_r3_u}) are
the Spherical Bessel functions, which is equivalent to having
\begin{eqnarray}
\nu(\kappa = 0) = \nu_0 = l_j + \frac{1}{2} \, .
\end{eqnarray}
If $\kappa$ is smaller than a critical value $\kappa_c$,
the shift can be written as
\begin{eqnarray}
\nu(\kappa < \kappa_c) = \nu_0 - \delta\,\nu \, ,
\end{eqnarray}
with $0 < \delta\,\nu  < \frac{1}{2}$.
For sufficiently small $\kappa$ the $\delta\,\nu$ correction 
is of order $|\kappa|$ ($\kappa^2$) for $l_j = 0$
($l_j \neq 0$), where the exact expression
can be consulted in Ref.~\cite{Birse:2005um}.
For larger values of $\kappa$, $\delta \nu$ is still a power series
in $\kappa$ and therefore perturbative.
However, this situation changes once $| \kappa |$ is large enough
for the shift to reach the value $\nu = l_j$:
above $\kappa_c$
the shift $\nu$ splits into two imaginary solutions of the type
\begin{eqnarray}
\nu(\kappa > \kappa_c) = l_j \pm i \rho(\kappa) \, .
\end{eqnarray}
Beyond this point, the renormalized angular momentum $\nu(\kappa)$
cannot be expressed as a power series in terms of $\kappa$,
as a consequence of the non-analytical character
of the split.
Consequently, we expect the standard perturbative treatment to fail
above this critical value $\kappa_c$.
The critical reduced momenta for the uncoupled channels can be consulted
in Table \ref{tab:kappa-c}.

\begin{table*}
\begin{center}
\begin{tabular}{|c|c||c|c||c|c||c|c|}
\hline \hline
$^3U_J$ & $\kappa_c$ & $^3C_J$ & $\kappa_c$ & $^5C_J$ & $\kappa_c$
& $^{1-5}C_J$ & $\kappa_c$ \\
\hline
$-$ & $-$ & $^3P_0$ & $1.259$ & $-$ & $-$ & $^1S_0$-${}^5D_0$ & $1.328$ \\
$^3P_1$ & $2.518$ & $^3S_1$-${}^3D_1$ & $1.367$ & $^5D_1$ & $8.333$
& $^1P_1$-$^5P_1$-$^5F_1$ & $4.314$ \\
$^3D_2$ & $8.333$ & $^3P_2$-${}^3F_2$ & $3.237$ & $^5P_2$-$^5F_2$ & $1.644$
& $^1D_2$-$^5S_2$-$^5D_2$-$^5G_2$ & $1.089$ \\
$^3F_3$ & $19.70$ & $^3D_3$-${}^3G_3$ & $7.913$ & $^5D_3$-$^5G_3$ & $5.594$
& $^1F_3$-$^5P_3$-$^5F_3$-$^5H_3$ & $2.615$ \\
\hline \hline
\end{tabular}
\end{center}
\caption{
Reduced critical momenta $\kappa_c$ for the different types of
coupled channels appearing in the ${\rm H\bar{H}}$ system
($j \leq 3$).
} \label{tab:kappa-c}
\end{table*}

\subsubsection{The Coupled Channel Case}

The (angular momentum) coupled channel case was worked out
in detail by Birse~\cite{Birse:2005um}.
The extension is trivial and only requires the following definitions.
In first place, the Schr\"odinger equation reads
\begin{eqnarray}
\label{eq:schroe_r3_c}
- {\bf u}_{k,j}'' + \left[ {\bf S}_j\,\frac{a_3}{r^3}
  + \frac{{\bf L}_j^2}{r^2} \right]\,{\bf u}_{k,j}(r) =
k^2 \, {\bf u}_{k,j}(r) \, ,
\end{eqnarray}
where there are $N$ angular momentum channels,
${\bf S}_j$ is the tensor matrix,
and ${\bf L}_j$ is a diagonal matrix containing
the value of the angular momenta
\begin{eqnarray}
{\bf L}_j^2 = {\rm diag}(l_1(l_1 + 1), \dots, l_N(l_N + 1)) \, .
\end{eqnarray}
As in the previous case, the general solution of the Schr\"odinger equation
is a linear combination of a the functions
\begin{eqnarray}
{\bf u}_{k,j}(r) = \sum_{\{l_j\}}\,
\left[ \alpha_{l_j} {\bm \xi}_{l_j}(r; k) + \beta_{l_j} {\bm \eta}_{l_j}(r; k)
\right] \, ,
\end{eqnarray}
where we sum over the possible values of the angular momenta.
In this case $\bm \xi$ and $\bm \eta$ are $N$-component vectors
that can be expressed as sums of Bessel functions
\begin{eqnarray}
{\bm \xi}_{l_j}(r; k) &=&
\sum_{m=-\infty}^{\infty} {\bm b}_m(\nu_{l_j})\,
\sqrt{r}\,J_{m + \nu_{l_j}}(k r) \, , \\
{\bm \eta}_{l_j}(r; k) &=&
\sum_{m=-\infty}^{\infty} (-1)^m {\bm b}_{-m}(\nu_{l_j})\,
\sqrt{r}\,J_{-m - \nu_{l_j}}(k r) \, , \nonumber \\
\end{eqnarray}
where we have labelled the different solutions by the subscript $_{l_j}$,
which indicates that for $\kappa = 0$ the solutions will behave
as free waves with angular momentum $l_j$.
The ${\bm b}_{m}(\nu_{l_j})$ coefficients are now vectors satisfying
a certain recursive relation, see Ref.~\cite{Birse:2005um}
for details.

Of course, what is important are the analytical properties of $\nu(\kappa)$.
The functional dependence $\nu = \nu(\kappa)$ is determined by
finding the zeros of 
\begin{eqnarray}
\label{eq:characteristic-c}
\det{\left( {\bf F}_j(\nu, \kappa) \right)} = 0 \, ,
\end{eqnarray}
where the characteristic function ${\bf F}_j(\nu, \kappa)$
is a ${N} \times {N}$ matrix.
The equation above admits $N$ solutions,
one for each value of the angular momentum,
which can be labelled by their value for $\kappa = 0$, that is
\begin{eqnarray}
\nu_{l_i}(\kappa = 0) = l_i + \frac{1}{2} \, ,
\end{eqnarray}
with $i = 1, \dots, N$.
We define ${\bf F}_j(\nu, \kappa)$ as follows
\begin{eqnarray}
{\bf F}_j(\nu, \kappa) \equiv {\bf f}_j(\nu) - \frac{\kappa^2}{\nu}\,
\left[ {\bf R}_0(\nu) - {\bf R}_0(-\nu)\right] \, ,
\end{eqnarray}
where ${\bf f}_j(\nu)$ is a diagonal matrix defined by
\begin{eqnarray}
{\bf f}_j(\nu) = {\rm diag}(
\nu^2 - (l_1 + \frac{1}{2})^2, \dots,
\nu^2 - (l_N + \frac{1}{2})^2) \, . \nonumber \\
\end{eqnarray}
In turn, ${\bf R}_0(\nu)$ can be expressed in terms of a recursive relation
\begin{eqnarray}
{\bf R}_n(\nu) = 
{\left[ {\bf f}_j(\nu) - {\bf S}_j\,\kappa^2\,{\bf S}_j\,{\bf R}_{n+1}(\nu)
\right]}^{-1 } \, . \nonumber \\
\end{eqnarray}
As in the uncoupled channel case, between 20 and 30 iterations are more
than enough to obtain a sufficiently accurate result.

The analytical properties of the shift $\nu_{l_i}(\kappa)$ are analogous
to what happened in the uncoupled channel case:
the shift begins its trajectory at $\nu_{l_i} = l_i + \frac{1}{2}$
for $\kappa = 0$ and moves slowly downwards.
Once the shift reaches the value $\nu_{l_i} = l_i$ at 
the critical value of the coupling $\kappa = \kappa_c$,
the shift splits into two complex conjugate solutions of the type
$\nu_{l_i}(\kappa) = l_i \pm i \rho_{l_i}(\kappa)$,
signalling the breakdown of a long range perturbative expansion
of the wave function.

In general, the first shift to split into the complex plane is the one
corresponding to the smallest angular momentum.
For that reason we will simply study the critical value of $\kappa$
corresponding to the lowest angular momentum in the coupled channel.
The results for the $^3C_J$, $^5C_J$ and $^{1-5}C_J$ families
of coupled channels with $j \leq 3$ are listed in Table \ref{tab:kappa-c}.
While the critical momenta for the $^3C_J$ type of
coupled channels was already determined by Birse in Ref.~\cite{Birse:2005um},
the results for the $^5C_J$ and $^{1-5}C_J$ coupled channels are new.

In general, for the $^3C_J$ channels we obtain values of $\kappa_c$
with are half the values obtained in Ref~\cite{Birse:2005um}.
The reason is simple:
the tensor operator matrix elements in the $^3C_J$ channels
for heavy meson molecules is half of the value of
the corresponding matrix in the two nucleon system.
In general,
the critical reduced momenta $\kappa_c$ grows rapidly with the angular momentum
of the system, with the exception of the $^3P_0$ partial wave for which
a relatively low value of $\kappa_c$ is obtained.
If we assume similar values of $a_3$, the tensor length scale,
the channel less likely to be perturbative is the 
$^1D_2$-$^5S_2$-$^5D_2$-$^5G_2$ partial wave,
corresponding to a $J^{PC} = 2^{++}$ $P^*\bar{P}^*$ meson molecule.

\subsection{Finite Pion Mass Effects}

In the previous paragraphs we have calculated the critical reduced momenta
$\kappa_c$ for a pure $1/r^3$ potential, corresponding to the form of OPE
for $\mu = 0$.
For $\mu > 0$ we expect the values of $\kappa_c$ to rise by a certain amount,
as the finite pion mass $e^{-\mu r}$ will decrease the effective strength
of the potential at large distances.
Curiously, we will see that the dependence of $\kappa_c$ on $\mu$
is actually triggered by the far from perfect separation of scales
in the chiral EFT, rather than from the finite pion mass alone.

The argument is the following:
we begin by solving the Schr\"odinger equation stepwise,
that is, we divide the possible radii $0 < r < \infty$ into several
regions of a given size $\Delta r$.
In particular, if we consider the region defined by the condition
\begin{eqnarray}
\label{def:region}
R - \frac{1}{2}\,\Delta r < r < R + \frac{1}{2}\,\Delta r \, ,
\end{eqnarray}
then, for $\mu \, \Delta r < 1$ we can approximate
the finite pion effects by the substitution
\begin{eqnarray}
\frac{a_3}{r^3}\,e^{-\mu r} \simeq \frac{a_3}{r^3}\,e^{-\mu R} \, ,
\end{eqnarray}
where we have ignored the $1/r$ and $1/r^2$ components of the tensor force
for simplicity~\footnote{These components can be implicitly taken into
account by a redefinition of $a_3$. However, provided $R$ is sufficiently
small, they are suppressed by a factor of $\mu\,R$.}.
At this point it is worth noticing that the explicit solutions of the
Schr\"odinger equation we have written in the preceding subsections
can also be applied in a specific subset of the real axis $0 < r < \infty$.
That is, if the $1/r^3$ potential is only valid
in the region defined by Eq.~(\ref{def:region}):
\begin{eqnarray}
V(r) = \frac{a_3}{r^3}
\,
\theta(|R + \frac{\Delta r}{2}| - r)\, 
\theta(r - |R - \frac{\Delta r}{2}|)
\, ,
\end{eqnarray}
then, Eqs.~(\ref{eq:uk_exp}), (\ref{eq:xi_exp}) and (\ref{eq:eta_exp}) 
represent the full solution of the Schr\"odinger equation in this region.
Moreover, the critical reduced momenta $\kappa_c$ calculated
for $0 < r < \infty$ are still valid
in $R - \Delta r / 2 < r < R + \Delta r / 2$.

The consequence for a decaying $e^{-\mu r} / r^3$ potential is that,
by making $\Delta r$ sufficiently small, we can define a
reduced critical momentum
\begin{eqnarray}
\kappa_c(\mu, R) = \kappa_c \, e^{\mu R} \, ,
\end{eqnarray}
which applies in the vicinity of $R$.
Therefore, the real $\kappa_c(\mu)$ setting the actual limits of perturbative
tensor OPE is the minimum of $\kappa_c(\mu, R)$,
which is in turn determined by the smallest radius
for which the OPE potential is valid.
In chiral EFT we expect this radius to be proportional
to the inverse of the breakdown scale,
$R_0 \propto 1/\Lambda_0$.
Naively, we expect $R_0$ to lie around $0.5\,{\rm fm}$, but
we cannot discard larger (or smaller) values.
In particular, a recent analysis of the convergence of the two-nucleon
potential in nuclear EFT~\cite{Baru:2012iv} suggests values around
$0.8\,{\rm fm}$ (and even higher).
Of course, the extrapolation of this latter value to the heavy meson-antimeson
potential is to be taken with a grain of salt, but we will nevertheless use
the $0.8\,{\rm fm}$ figure only as an upper bound for $R_0$.
For $\mu = m_{\pi}$, the $R_0 = 0.5-0.8\,{\rm fm}$ window yields
\begin{eqnarray}
\label{eq:kappa_pi}
\kappa(m_{\pi}) = \kappa_c(m_{\pi}, R_0) \simeq
(\sqrt{2} - \sqrt{3})\,\kappa_c \, ,
\end{eqnarray}
where we have approximated $e^{m_{\pi} R_0}$ by $\sqrt{2}$ ($\sqrt{3}$)
for $R_0 = 0.5\,(0.8)\,{\rm fm}$.
In the nucleon-nucleon case the $\sqrt{2}$ factor approximately corresponds
to the observed mismatch between the critical momentum obtained
by Birse for the $^3S_1$-${}^3D_1$ channel
($p_c = 66\,{\rm MeV}$)~\cite{Birse:2005um}
and the momentum at which the FMS calculations~\cite{Fleming:1999ee}
fails ($p_c \sim 100\,{\rm MeV}$).
However, a more recent formulation of nuclear EFT
with perturbative OPE~\cite{Beane:2008bt}
suggests a larger $p_c \geq 150\,{\rm MeV}$,
requiring $R_0 \geq 1.0\,{\rm fm}$ at least.

\section{Discussion and Conclusions}
\label{sec:dis}

\begin{table*}
\begin{center}
\begin{tabular}{|c|c|c|c|c|c|c|}
\hline \hline
$I^G(J^{PC})$ & $^{2S+1}L_J$ & $\rm H\bar{H}$
& $p_{\rm crit}(0)$ & $B_{\rm crit}(0)$ & $p_{\rm crit}(\mu)$ & $B_{\rm crit}(\mu)$ 
\\
\hline \hline
$0^{\pm}(1^{+\pm})$ & $^3S_1$-$^3D_1$ & $\rm D\bar{D}^*$
& $290^{+120}_{-80}$ & $42^{+46}_{-19}$ 
& $290^{+120}_{-80}$ & $42^{+46}_{-19}$ 
\\
\hline
$0^{+}(0^{++})$ & $^1S_0$-$^5D_0$ & $\rm D^*\bar{D}^*$
& $270^{+120}_{-70}$ & $36^{+38}_{-17}$
&  $420^{+190}_{-120}$ & $89^{+97}_{-41}$ \\
$0^{-}(1^{+-})$ & $^3S_1$-$^3D_1$ & $\rm D^*\bar{D}^*$
& $280^{+120}_{-80}$ & $38^{+41}_{-18}$ 
& $440^{+200}_{-120}$ & $90^{+100}_{-50}$ \\
$0^{+}(2^{++})$ & $^1D_2$-$^5S_2$-$^5D_2$-$^5G_2$ & $\rm D^*\bar{D}^*$
& $220^{+100}_{-60}$ & $24^{+26}_{-11}$
& $350^{+160}_{-100}$ & $60^{+65}_{-30}$ \\
\hline \hline
$1^{\mp}(1^{+\pm})$ & $^3S_1$-$^3D_1$ & $\rm B\bar{B}^*$
& $450^{+260}_{-140}$ & $38^{+56}_{-19}$
& $690^{+400}_{-220}$ & $90^{+130}_{-50}$ \\
\hline
$1^{-}(0^{++})$ & $^1S_0$-$^5D_0$ & $\rm B^*\bar{B}^*$
& $440^{+240}_{-140}$ & $36^{+51}_{-19}$
& $690^{+390}_{-220}$ & $90^{+130}_{-50}$ \\
$1^{+}(1^{+-})$ & $^3S_1$-$^3D_1$ & $\rm B^*\bar{B}^*$
& $450^{+250}_{-140}$ & $38^{+55}_{-20}$
& $710^{+410}_{-230}$ & $90^{+140}_{-50}$ \\
$1^{-}(2^{++})$ & $^1D_2$-$^5S_2$-$^5D_2$-$^5G_2$ & $\rm B^*\bar{B}^*$
& $360^{+200}_{-110}$ & $24^{+35}_{-12}$
& $560^{+320}_{-170}$ & $60^{+87}_{-33}$ \\
\hline \hline
\end{tabular}
\end{center}
\caption{
Critical value of the momenta for the tensor piece of OPE
for the $s$-wave isoscalar $D\bar{D}^*$ and $D^*\bar{D}^*$
and isovector $B\bar{B}^*$ and $B^*\bar{B}^*$ systems.
We compute the critical value both in the $\mu = 0$ limit
and in the finite $\mu$ limit.
In the second case, we usually have $\mu = m_{\pi}$,
with the exception of the $D\bar{D}^*$ ($B \bar{B}^*$) case
in which $|\mu| = 31\,{\rm MeV}$ ($\mu = 131\,{\rm MeV}$),
generating no significant deviation from $\mu = 0$ ($\mu = m_{\pi}$).
The respective value of the critical momenta and maximum binding energies
for the isovector $D\bar{D}^*$ and $D^*\bar{D}^*$ molecules is
three (nine) times larger than for their isoscalar counterparts.
The contrary happens with the isoscalar $B\bar{B}^*$ and $B^*\bar{B}^*$
molecules, for which the critical momenta are $1/3$ of the isovector
case and the maximum binding energies $1/9$.
} \label{tab:charm-bottom}
\end{table*}

The critical values $\kappa_c$ can be converted into critical momenta
by dividing by the tensor length scale $a_3$
\begin{eqnarray}
a_3 = |\tau|\,\frac{\mu_{\rm H\bar{H}} g^2}{4 \pi f_{\pi}^2} \, .
\end{eqnarray}
In the equation above, $\tau = 2 I(I+1) - 3$ is the eigenvalue of the isospin
operator $\vec{\tau}_1 \cdot \vec{\tau}_2$ and $\mu_{\rm H\bar{H}}$
the reduced mass of the heavy meson molecule under consideration.
For the axial coupling constant $g$, we take $g = 0.6 \pm 0.1$
in the charm sector ($\rm D^{(*)}\bar{D}^{(*)}$)
and $g = 0.5 \pm 0.1$ in the bottom one
($\rm B^{(*)}\bar{B}^{(*)}$).
For the charm mesons, the value of $g$ is known from the $D^* \to D\pi$
and $D^* \to D\gamma$ decays~\cite{Ahmed:2001xc,Anastassov:2001cw}.
This yields $g = 0.59\, \pm 0.01\, \pm 0.07$,
which we round to $g = 0.6\, \pm 0.1$.
For the bottom mesons, $g$ is experimentally unknown,
even though from heavy quark symmetry we should expect a value similar
to the charm mesons one.
There are however theoretical determinations, for example
from lattice QCD~\cite{Abada:2003un,Negishi:2006sc,Detmold:2007wk,
Becirevic:2009yb} or from the QCD Dyson-Schwinger
equations~\cite{ElBennich:2010ha}
to name just a few of them,
usually ranging from $0.3$ to $0.7$~\cite{Li:2010rh}.
The value we have chosen, $g = 0.5 \pm 0.1$, represent a compromise
between the previous determinations, but it should be kept in mind
that the uncertainties for the bottom sector may be much larger.

The critical momentum (including finite pion mass corrections) is defined as
\begin{eqnarray}
p_{\rm crit}(\mu) = \frac{\kappa_c}{a_3}\,e^{+\mu R_0} \, ,
\end{eqnarray}
with $R_0 = 0.65 \pm 0.15\,{\rm fm}$ (i.e. $R_0 = 0.5-0.8\,{\rm fm}$),
and $\kappa_c$ to be taken from Table \ref{tab:kappa-c}.
We can also translate $p_{\rm crit}(\mu)$ into critical binding energies by
\begin{eqnarray}
B_{\rm crit}(\mu) = \frac{p_{\rm crit}^2(\mu)}{2 \mu_{\rm {H\bar H}}} \, ,
\end{eqnarray}
where we can make the approximation
$B_{\rm crit}(m_{\pi}) \simeq (2.5 \pm 0.5) B_{\rm crit}(0)$.
It is also interesting that the critical binding energy scales as $1/g^4$.
This dependence will generate remarkable uncertainties on the value of
$B_{\rm crit}$ for bottom meson molecules.

In the charm isovector sector, that is, the $\rm D^{(*)}\bar{D}^{(*)}$
molecules with $I = 1$ and $\tau = 1$,
the critical momenta are of the order of $0.8-1.1\,{\rm GeV}$,
similar to the chiral breakdown scale $\Lambda_0 \sim 0.5-1.0\,{\rm GeV}$.
In this case, central and tensor OPE are expected to be a perturbation
in all the range of validity of the EFT.
On the other extreme we have the bottom isoscalar sector,
for which the critical momenta are of the order of $m_{\pi}$
($\sim 200\,{\rm MeV}$ for $\mu = m_{\pi}$).
For these systems the onset of non-perturbative OPE start
at around $4\,{\rm MeV}$ in the chiral limit,
or around $10\,{\rm MeV}$ if we take into account
the finite pion mass effects.
It should be noted that the previous limits could be more stringent
if $g$ is larger than the expected $0.5$, as the critical binding
depends on $1 / g^4$.
However, the bottomline is that in these heavy meson molecules
OPE is expected to be non-perturbative unless the bound states
are genuinely shallow.
The corresponding EFT treatment is the one employed
in Ref.~\cite{Nieves:2011zz} for exploring
prospective $\rm B^* \bar B$ states.

The two interesting cases are
the isovector charm and isoscalar bottom meson molecules:
in these two systems the values of the critical momenta lie between
$m_{\pi}$ and $\Lambda_0$, defining a specific window in which OPE
is perturbative.
The critical momenta for the charm and bottom sectors can be found in Table
\ref{tab:charm-bottom}.
For the charm isoscalar the critical binding energy lies around
$25-40\,{\rm MeV}$ ($60-90\,{\rm MeV}$) for $\mu = 0$ ($\mu = m_{\pi}$).
In this regard, the $\rm D^*\bar{D}$/$\rm D\bar{D}^*$ systems
are specially interesting in the sense that they naturally
fulfill the condition $|\mu| \sim 0$~\footnote{Strictly speaking
$|\mu| = 31\,{\rm MeV} \ll m_{\pi}$, which is small enough as
to consider $\mu \sim 0$ as a pretty good approximation.}.
This means that the $X(3872)$ is well within the energy range for which
the ${\rm LO}$ EFT description consists only on contact range interactions.
In particular, the critical momentum of $290\,{\rm MeV}$ implies that
the charged $\rm D^{*+} D^{-}$/$\rm D^{*-} D^{+}$ component of
the $X(3872)$ can also be accounted for without the explicit
inclusion of pion exchanges at the lowest order.
This represents a remarkable simplification that allows, for example,
to reinterpret the explanation of
Gamermann et al.~\cite{Gamermann:2009fv,Gamermann:2009uq}
for the branching ratio
\begin{eqnarray}
\frac{{\Gamma}\left[X(3872) \to J/\psi \pi^{+} \pi^{-} \pi^0\right]}
{{\Gamma}\left[X(3872) \to J/\psi \pi^{+} \pi^{-}\right]} \, ,
\end{eqnarray}
as a result within a specific EFT framework,
or to extend X-EFT~\cite{Fleming:2007rp}
to the charged channels of the $X(3872)$.

In the isovector bottom sector, to which the recently discovered
$Z_b(10610)$ and $Z_b(10650)$ belong, the critical energies range
from $60$ to $90\,{\rm MeV}$, although they greatly depend
on the axial coupling $g$, which value is not that well-known.
This energy is in fact of the order of the splitting between the different
channels ($\Delta_B = 46\,{\rm MeV}$/$92\,{\rm MeV}$),
indicating that non-perturbative OPE is probably not needed
until we reach binding energies such that
coupled channel effects cannot be longer ignored.
In this regard, the pionless EFT exploration of Mehen and
Powell~\cite{Mehen:2011yh} probably provides a perfectly
good ${\rm LO}$ description of the prospective HQSS partners of the $Z_b$'s
as far as their binding energies is
less than $40-50\,{\rm MeV}$.
Moreover, the inclusion of perturbative pions in the subleading orders of
the aforementioned EFT will be a direct extension of X-EFT~\cite{Fleming:2007rp}
to the bottom meson-antimeson system.
We can call this theory Z-EFT.

Of course, the previous results are to be held in proper context:
they refer to the EFT framework rather than to a more
fundamental or phenomenological level of description
of heavy meson molecules.
In phenomenological approaches non-perturbative pion exchanges and
coupled channel effects may be important in suggesting the possibility
or explaining the location
of heavy molecular states~\cite{Sun:2011uh,Ohkoda:2011vj},
in particular if the states are shallow,
in which case a fine tuning takes place between the different long and
short range effects that are explicitly included.
In contrast, in the EFT approach these fine tunings are absorbed
in the counterterms, which are set to reproduce the already
known position of a certain state.
This in turn stabilizes the calculations and reduces the impact of
new physical effects we may subsequently incorporate in the EFT
description as they are expected to be constrained by power counting.
The price to pay is the inability of the EFT framework to predict
the positions of these states from first principles.

To summarize, we have found that pion exchange effects are perturbative
in most types of heavy meson molecules over the expected range
of applicability of heavy meson EFT.
The exception is the isoscalar bottom meson molecules, for which OPE
is particularly strong and presumably non-perturbative.
To a lesser extent, in the isoscalar charm sector
non-perturbative OPE may be required
if the molecular states are deep enough (around $80\,{\rm MeV}$ or more).
At lowest order, below the critical binding energies we have computed,
the EFT description is a contact-range theory.
Surprisingly, this description holds for momenta much larger
than naively expected.

\begin{acknowledgments}

I would like to thank E. Ruiz Arriola for discussions
and a critical and careful reading of the manuscript
and J. Nieves for discussions.
This work was supported by the DGI under contract FIS2011-28853-C02-02,
the Generalitat Valenciana contract PROMETEO/2009/0090,
the Spanish Ingenio-Consolider 2010 Program CPAN
(CSD2007-00042) and the EU Research Infrastructure Integrating Initiative
HadronPhysics2.

\end{acknowledgments}

\appendix
\section{Derivation of The Leading Order Potential in Heavy Hadron Chiral Perturbation Theory}
\label{app:LO-pot}

In this appendix, we derive the ${\rm LO}$ potential between
a heavy mesons and anti-meson within heavy hadron chiral
perturbation theory (HHChPT)~\cite{Wise:1992hn}.
The ${\rm LO}$ potential can be decomposed into a contact-range
and a finite-range piece.
The later can be identified with the well-known one pion exchange (OPE)
potential.
We also compute the complete partial wave projection of the ${\rm LO}$
potential.

\subsection{The Effective Lagrangian at Leading Order}

We define the heavy meson (antimeson) fields in terms of
the pseudoscalar and vector meson (antimeson) fields
in the following way.
\begin{eqnarray}
H^{Q} &=& \frac{1 + \slashed v}{2}\,
\left[  P^{* \mu}_v \gamma^{\mu} - P_v \gamma^5 \right] \, , \\
\bar{H}^{Q} &=& \gamma_0 {H^{Q}\,}^{\dagger} \gamma_0 \, , \\
\nonumber \\
H^{\bar{Q}} &=& 
\left[  \bar{P}^{* \mu}_v \gamma^{\mu} - \bar{P}_v \gamma^5 \right] \,
\frac{1 - \slashed v}{2} \, , \\
\bar{H}^{\bar{Q}} &=& \gamma_0 {H^{\bar{Q}}\,}^{\dagger} \gamma_0 \, .
\end{eqnarray}
where $\gamma_{\mu}$ and $\gamma_5$ are the Dirac gamma matrices,
$v$ is the velocity parameter and $\slashed v = v^{\mu}  \gamma_{\mu}$.
The field $P_v^{(*\mu)}$ ($\bar{P}_v^{(*\mu)}$)
destroys a heavy meson (antimeson) of velocity $v$
and polarization $\mu$ for the vector meson.
The subscript $_v$ also indicates that the normalization of
the one-particle states is given by
\begin{eqnarray}
\langle P_v(\vec{k}) | P_{v'}(\vec{k}') \rangle &=& 2 v^0 \,\delta_{vv'}\,(2\pi)^3\,\delta^{(3)}(\vec{k} - \vec{k}') \, , \\
\langle P^{*\mu}_v(\vec{k}) | P^{*\nu}_{v'}(\vec{k}') \rangle &=& 2 v^0 \,\delta_{vv'}\,(2\pi)^3\,\delta^{(3)}(\vec{k} - \vec{k}')\,\delta_{\mu \nu} \, ,
\nonumber \\
\end{eqnarray}
depending on whether we are considering the pseudoscalar or vector meson case.

The HHChPT Lagrangian at lowest order ($Q^0$) can be decomposed
as the sum of two different contributions:
\begin{eqnarray}
\mathcal{L}^{(0)} =
\mathcal{L}^{(0)}_{\rm 4 H} + \mathcal{L}^{(0)}_{\pi {\rm H H}} \, ,
\end{eqnarray}
where the first term represents a 4-meson interaction vertex
and the second the meson-pion vertex.
The contact interaction term reads~\cite{AlFiky:2005jd}
\begin{eqnarray}
\mathcal{L}_{\rm 4H}^{(0)} &=&
D_{0a}\,{\rm Tr}\left[ \bar{H}^{Q} H^{Q} \gamma_{\mu} \right]
\,{\rm Tr}\left[  H^{\bar{Q}}\bar{H}^{\bar{Q}} \gamma^{\mu} \right]
\nonumber \\
&+& D_{0b}\,{\rm Tr}\left[ \bar{H}^{Q} H^{Q} \gamma_{\mu} \gamma_5  \right] 
\,{\rm Tr}\left[  H^{\bar{Q}}\bar{H}^{\bar{Q}} \gamma^{\mu} \gamma_5 \right]
\nonumber \\
&+& E_{0a}\,{\rm Tr}\left[ \bar{H}^{Q} \tau_i H^{Q} \gamma_{\mu} \right]
\,{\rm Tr}\left[  H^{\bar{Q}} \tau_i \bar{H}^{\bar{Q}} \gamma^{\mu} \right]
\nonumber \\
&+& E_{0b} \,{\rm Tr}\left[ \bar{H}^{Q} \tau_i H^{Q} \gamma_{\mu} \gamma_5 
\right] 
\,{\rm Tr}\left[  H^{\bar{Q}}\bar{H}^{\bar{Q}} \tau_i \gamma^{\mu} \gamma_5 \right] \, ,
\nonumber \\ \label{eq:Lagrangian_LO_contact}
\end{eqnarray}
where $\tau_i$ are the isospin matrices (i.e. the Pauli matrices).
That is, in the previous Lagrangian $D_{0a}$ and $D_{0b}$
are isospin independent, while $E_{0a}$ and $E_{0b}$ are isospin dependent.
In general we will be only considering a specific isospin channel,
in which case we will generically define
\begin{eqnarray}
C_{0a} &=& D_{0a} + \vec{\tau}_1 \cdot \vec{\tau}_2 \, E_{0a} \, , \\
C_{0b} &=& D_{0b} + \vec{\tau}_1 \cdot \vec{\tau}_2 \, E_{0b} \, .
\end{eqnarray}
For the Lagrangian describing the pion-meson vertex we have~\cite{Wise:1992hn}
\begin{eqnarray}
\mathcal{L}_{\pi H \bar{H}}^{(0)} &=& i\,\frac{g}{2}\,{\rm Tr}
\left[
\bar{H}^{Q} H^{Q} \gamma_{\mu} \gamma_5 \,
(\xi^{\dagger} \partial^{\mu} \xi - \xi \partial^{\mu} \xi^{\dagger})
\right] \nonumber \\
&-& i\,\frac{g}{2}\,{\rm Tr}
\left[
{H}^{\bar Q} {\bar H}^{\bar Q} \gamma_{\mu} \gamma_5 \,
(\xi^{\dagger} \partial^{\mu} \xi - \xi \partial^{\mu} \xi^{\dagger})
\right] \, , \nonumber \\ \label{eq:Lagrangian_LO_OPE}
\end{eqnarray}
where $\xi$ is defined as
\begin{eqnarray}
\xi &=& e^{\frac{i}{f_{\pi}} M} \, , \\
M &=&
\begin{pmatrix}
\frac{\pi^0}{\sqrt{2}} & \pi^{+} \\
\pi^{-} & -\frac{\pi^0}{\sqrt{2}}
\end{pmatrix}
 = \frac{1}{\sqrt{2}}\,\vec{\tau} \cdot \vec{\pi} \, ,
\end{eqnarray}
with $\pi$ the pion fields and $f_{\pi} = 132\,{\rm MeV}$.
The isospin indices convention for the heavy meson and antimeson fields is
\begin{eqnarray}
P_v^{(*\mu)} &=& 
\begin{pmatrix}
P_{v,+}^{(*\mu)} \\
P_{v,0}^{(*\mu)}
\end{pmatrix} \quad \mbox{and} \quad
\bar{P}_v^{(*\mu)} =
\begin{pmatrix}
-\bar{P}_{v,0}^{(*\mu)} \\
\phantom{+}P_{v,-}^{(*\mu)}
\end{pmatrix} \, . \nonumber \\
\end{eqnarray}
This convention explains the minus sign
in the second line of Eq.~(\ref{eq:Lagrangian_LO_OPE}).
That is, we have performed a G-parity transformation to obtain the chiral
Lagrangian for the heavy anti-meson fields from the heavy meson one.

\subsection{The Non-Relativistic Normalization}

In the heavy quark limit, $m_Q \to \infty$, the static potential between
two heavy mesons is a well-defined object.
Relativistic effects are suppressed, and the two heavy meson system
can be effectively described in terms of
non-relativistic quantum mechanics.
For making the non-relativistic transition we specify the velocity parameter
to be $v = (1, \vec{0})$ and employ a new normalization of
the heavy meson fields defined by
\begin{eqnarray}
P^{(*j)} = {\sqrt{2}} P_v^{(*j)} \, ,
\end{eqnarray}
where we have substituted the Greek index $\mu$ by the latin index $j$,
as the $\mu = 0$ polarization component is completely irrelevant
for $v = (1, \vec{0})$.
In this heavy meson field normalization, the one-particle states have the
usual non-relativistic normalization
\begin{eqnarray}
\langle P(\vec{k}) | P(\vec{k}') \rangle &=& (2\pi)^3\,\delta^{(3)}(\vec{k} - \vec{k}') \, , \\
\langle P^{*i}(\vec{k}) | P^{*j}(\vec{k}') \rangle &=& (2\pi)^3\,\delta^{(3)}(\vec{k} - \vec{k}')\,\delta_{i j} \, ,
\end{eqnarray}
which is more convenient for the definition of
the quantum mechanical potential.

\subsection{The Contact Range Potential}

In the non-relativistic normalization,
after expanding the ${\rm H^{(Q)}}$ and ${\rm \bar H}^{(Q)}$ fields,
the contact range Lagrangian can be rewritten as
\begin{eqnarray}
\mathcal{L}_{\rm 4H \, (a)} &=& - C_{0a} 
\,({P}^{\dagger} P + {P}^{*\,\dagger} P^*) \,
(\bar{P} {\bar P}^{\dagger} +  \bar{P}^* \bar{P}^{*\,\dagger}) \, \nonumber \\
&& + C_{0b}\,
(P^*\,P^{\dagger} + P\,P^{*\,\dagger})\,
({\bar P}^*\,{\bar P}^{\dagger} + {\bar P}\,{\bar P}^{*\,\dagger}) \nonumber \\
&& - i\,C_{0b} \Big[ (P^*\,P^{\dagger} + P\,P^{*\,\dagger})\,
({\bar P}^{*\,\dagger} \times {\bar P}^*) \nonumber \\
&& \phantom{C2 i} - ({P}^{*\,\dagger} \times {P}^*)\,
({\bar P}^*\,{\bar P}^{\dagger} + {\bar P}\,{\bar P}^{*\,\dagger})
\Big] \nonumber \\
&& + C_{0b}\,({P}^{*\,\dagger} \times {P}^*)\,({\bar P}^{*\,\dagger}
\times {\bar P}^*) \, .
\end{eqnarray}
In the expressions above, the polarization of the vector mesons
has been made implicit to simplify the notation.
The potential corresponding to the Lagrangian above
can mix different particle channels.
Thus, it is convenient to write the potential in the coupled channel basis
\begin{eqnarray}
{\mathcal B}_{\rm H\bar{H}} = \left\{
| P\bar{P} \rangle, | P^*\bar{P} \rangle,  | P\bar{P}^* \rangle, 
| P^*\bar{P}^* \rangle \right\} \, ,
\end{eqnarray}
for which we have (in momentum space)
\begin{widetext}
\begin{eqnarray}
V_C(\vec{q}) &=&
C_{0a}\,\begin{pmatrix}
1 & 0 & 0 & 0 \\
0 &  \vec{\epsilon_{1'}}^* \cdot \vec{\epsilon_{1}} & 0 & 0 \\
0 & 0 & \vec{\epsilon_{2'}}^* \cdot \vec{\epsilon_{2}} & 0 \\
0 & 0 & 0 &
\vec{\epsilon_{1'}}^* \cdot \vec{\epsilon_{1}} \, 
\vec{\epsilon_{2'}}^* \cdot \vec{\epsilon_{2}}
\end{pmatrix} 
+ C_{0b}\,\begin{pmatrix}
0 & 0 & 0 &  -\vec{\epsilon_{1'}}^* \cdot \vec{\epsilon_{2'}}^* \\
0 & 0 & -\vec{\epsilon_{1}} \cdot \vec{\epsilon_{2'}}^* & 
+\vec{S}_1 \cdot \vec{\epsilon_{2'}}^*  \\
0 & - \vec{\epsilon_{1'}}^* \cdot \vec{\epsilon_{2}} & 0 & -\vec{\epsilon_{1'}}^* \cdot \vec{S}_2 \\
-\vec{\epsilon_{1}} \cdot \vec{\epsilon_{2}} & 
+\vec{S}_1 \cdot \vec{\epsilon_{2}} & 
-\vec{\epsilon_{1}} \cdot \vec{S}_2 &
+\vec{S}_1 \cdot \vec{S}_2
\end{pmatrix} \, . \nonumber \\
\end{eqnarray}
\end{widetext}
In the equation above, $\vec{\epsilon}_{1(2)}$/$\vec{\epsilon}_{1'(2')}$
represents the polarization of the incoming particles $1(2)$
or the outgoing particles $1'(2')$.
In turn, the matrix elements of the spin-1 operator $\vec{S}_{1(2)}$
are equivalent to the vector product of the polarization wave functions,
that is
\begin{eqnarray}
i\,\langle 1 \lambda' | \vec{S} | 1 \lambda \rangle =
(\vec{\epsilon_{\lambda'}}^* \times \vec{\epsilon_{\lambda}}) \, ,
\end{eqnarray}
where $| 1 \lambda^{(')} \rangle$ is the vector corresponding 
to the polarization wave function ${\vec{\epsilon}_{\lambda^{(')}}}$.

\subsection{The Finite Range OPE Potential}

For obtaining the finite range piece of the potential,
we start by considering the ${\rm HH\pi}$ vertex in
the non-relativistic normalization,
for which the Lagrangian reads
\begin{eqnarray}
\mathcal{L}_{\pi H H} &=& \frac{g}{\sqrt{2} f_{\pi}}
\left[ (P^* P^{\dagger} + P^{*\,\dagger} P) + i\,(P^{*\dagger} \! \times \! P^*)
\right] \vec{\tau} \cdot \partial \vec{\pi}
\nonumber \\
&+& \frac{g}{\sqrt{2} f_{\pi}}
\left[ (\bar{P}^* \bar{P}^{\dagger} + {\bar P}^{*\,\dagger} {\bar P})
- i\,({\bar P}^{*\dagger} \! \times \! {\bar P}^*)
\right] \vec{\tau} \cdot \partial \vec{\pi} \, , \nonumber \\
\end{eqnarray}
where the polarization indices are implicit.
If we define the non-relativistic amplitude as follows
\begin{eqnarray}
\mathcal{A}(H \to H\pi^{a}) =
i\,\langle H | \mathcal{L}_{\pi HH} | H \pi^a \rangle \, ,
\end{eqnarray}
we find
\begin{eqnarray}
{\mathcal A}(P \to P \pi^a) &=& 0 \, , \\
{\mathcal A}(P^{*i} \to P \pi^a) &=& \frac{g}{f_{\pi}}\,
\frac{\tau^a}{\sqrt{2}}\, \vec{\epsilon}_{i} \cdot \vec{q} \, , \\
{\mathcal A}(P^{*i} \to P^{*j} \pi^a) &=& -i\,\frac{g}{f_{\pi}}\,
\frac{\tau^a}{\sqrt{2}}\, 
\vec{S} \cdot \vec{q} \, ,
\end{eqnarray}
where $\vec{q}$ is the momentum of the outgoing pion, $a$ the isospin
index of the pion and $\vec{\epsilon}_{i}$ the polarization
wave function of the heavy vector meson $P^{*i}$.
The amplitudes with heavy anti-mesons can be obtained either
from the Lagrangian or from a G-parity transformation of
the expressions above, yielding
\begin{eqnarray}
{\mathcal A}(\bar{P} \to \bar{P} \pi^a) &=& 0 \, , \\
{\mathcal A}(\bar{P}^{*i} \to \bar{P} \pi^a) &=& \frac{g}{f_{\pi}}\,
\frac{\tau^a}{\sqrt{2}}\, \vec{\epsilon}_{i} \cdot \vec{q} \, , \\
{\mathcal A}(\bar{P}^{*i} \to \bar{P}^{*j} \pi^a) &=& i\,\frac{g}{f_{\pi}}\,
\frac{\tau^a}{\sqrt{2}}\, 
\vec{S} \cdot \vec{q} \, ,
\end{eqnarray}
where only the $\rm {\bar P}^* {\bar P}^* \pi$ vertex changes sign.

From the non-relativistic amplitudes above, the OPE potential
in the heavy quark limit is given by
\begin{widetext}
\begin{eqnarray}
V^{(0)}_{F, \rm H\bar{H}}(\vec{q}) &=&
\frac{g^2}{2 f_{\pi}^2}\,\vec{\tau}_1 \cdot \vec{\tau}_2
\,\frac{1}{{\vec{q}\,}^2 + m_{\pi}^2}\,
\,\begin{pmatrix}
0 & 0 & 0 & 
-\vec{\epsilon_{1'}}^* \cdot \vec{q} \, \vec{\epsilon_{2'}}^* \cdot \vec{q} \\
0 & 0 &
-\vec{\epsilon_{1}} \cdot \vec{q} \, \vec{\epsilon_{2'}}^* \cdot \vec{q} & 
+\vec{S}_1 \cdot \vec{q} \, \vec{\epsilon_{2'}}^* \cdot \vec{q} \\
0 & - \vec{\epsilon_{1'}}^* \cdot \vec{q} \, \vec{\epsilon_{2}} \cdot \vec{q}
& 0 & - \vec{\epsilon_{1'}}^* \cdot \vec{q} \, \vec{S}_2 \cdot \vec{q} \\
-\vec{\epsilon_{1}} \cdot \vec{q} \, \vec{\epsilon_{2}} \cdot \vec{q} & 
+\vec{S}_1 \cdot \vec{q} \, \vec{\epsilon_{2}} \cdot \vec{q} & 
-\vec{\epsilon_{1}} \cdot \vec{q} \, \vec{S}_2 \cdot \vec{q} &
+\vec{S}_1 \cdot \vec{q} \, \vec{S}_2 \cdot \vec{q}
\end{pmatrix} + \mathcal{O}(\frac{1}{m_Q})
\, , \nonumber \\
\end{eqnarray}
\end{widetext}
where we have already particularized for the heavy meson-antimeson case.
If we consider the effect of the splitting between the masses of the
${\rm P\bar{P}}$, ${\rm P\bar{P}^*}$/${\rm P^*\bar{P}}$,
and ${\rm P^*\bar{P}^*}$ systems,
then we should substitute the pion propagator by
\begin{eqnarray}
\frac{1}{{\vec{q}\,}^2 + m_{\pi}^2} \to \frac{1}{{\vec{q}\,}^2 + \mu^2}
\end{eqnarray}
in the non-diagonal terms of the potential matrix, where the effective
pion mass is given by $\mu^2 = m_{\pi}^2 - \Delta_Q^2$,
with $\Delta_Q$ the energy splitting.
This corrections are fundamental in the charm sector,
where $\Delta_Q$ is of the order of $m_{\pi}$.
A further problem arises in the ${\rm P\bar{P}} \to {\rm P^*\bar{P}^*}$
and ${\rm P\bar{P}^*}$/${\rm P^*\bar{P}} \to {\rm P^*\bar{P}^*}$
transitions, where either the initial or final heavy meson states
will be off-shell.
In this case it is not entirely clear whether the static approximation
can be employed to define a transition potential between the different
heavy meson states in the charm sector.
Luckily this problem is not present in the heavy quark limit, in which
the possible off-shell effects are strongly suppressed.

We can calculate the coordinate space representation of the OPE potential
by Fourier transforming the momentum space potential
\begin{eqnarray}
V^{(0)}_{F, \rm H \bar H}(\vec{r}) &=&
\int\,\frac{d^3\,q}{(2 \pi)^3}\,{V}^{(0)}_{F, \rm H \bar H}(\vec{q})
\,e^{-i \vec{q} \cdot \vec{r}} \, ,
\end{eqnarray}
from which we obtain
\begin{eqnarray}
V^{(0)}_{F, {\rm H \bar H}}(\vec{r}) &=&
\vec{\tau}_2 \cdot \vec{\tau}_1\,{\bf C}_{12}(\hat{r})\,\frac{g^2}{6 f_{\pi}^2}
\,\delta(\vec{r}) \nonumber \\ &-& 
\vec{\tau}_2 \cdot \vec{\tau}_1\,\Big[
{\bf C}_{12}(\hat{r}) W_C (r) + {\bf S}_{12}(\hat{r}) \, W_T (r) \Big]
\nonumber \\
&+& \mathcal{O}(\frac{1}{m_Q}) \, ,
\end{eqnarray}
with $W_C(r)$ and $W_T(r)$ defined in Eqs.~(\ref{eq:WC}) and (\ref{eq:WC}).
The central and tensor operators ${\bf C}_{12}(\hat{r})$ and
${\bf S}_{12}(\hat{r})$ are $4 \times 4$ matrices
in the particle channel basis ${\mathcal B}_{\rm H\bar{H}}$.
Their explicit representation is
\begin{widetext}
\begin{eqnarray}
{\bf C}_{12}(\hat{r}) &=&
\begin{pmatrix}
0 & 0 & 0 &  -\vec{\epsilon_{1'}}^* \cdot \vec{\epsilon_{2'}}^* \\
0 & 0 & -\vec{\epsilon_{1}} \cdot \vec{\epsilon_{2'}}^* & 
+\vec{S}_1 \cdot \vec{\epsilon_{2'}}^*  \\
0 & - \vec{\epsilon_{1'}}^* \cdot \vec{\epsilon_{2}} & 0 & -\vec{\epsilon_{1'}}^* \cdot \vec{S}_2 \\
-\vec{\epsilon_{1}} \cdot \vec{\epsilon_{2}} & 
+\vec{S}_1 \cdot \vec{\epsilon_{2}} & 
-\vec{\epsilon_{1}} \cdot \vec{S}_2 &
+\vec{S}_1 \cdot \vec{S}_2
\end{pmatrix} \, , \\
{\bf S}_{12}(\hat{r}) &=&
\,\begin{pmatrix}
0 & 0 & 0 & 
- S_{12}(\vec{\epsilon_{1'}}^*, \vec{\epsilon_{2'}}^*, \hat{r}) \\
0 & 0 &
- S_{12}(\vec{\epsilon_{1}}, \vec{\epsilon_{2'}}^*,  \hat{r}) & 
+ S_{12}(\vec{S}_1 , \vec{\epsilon_{2'}}^*, \hat{r}) \\
0 & - S_{12}(\vec{\epsilon_{1'}}^* , \vec{\epsilon_{2}} , \hat{r})
& 0 & - S_{12}(\vec{\epsilon_{1'}}^* , \vec{S}_2 , \hat{r}) \\
- S_{12} (\vec{\epsilon_{1}} , \vec{\epsilon_{2}} , \hat{r}) & 
+ S_{12} (\vec{S}_1 , \vec{\epsilon_{2}} , \hat{r}) & 
- S_{12} (\vec{\epsilon_{1}} , \vec{S}_2 , \hat{r}) &
+ S_{12} (\vec{S}_1 , \vec{S}_2 , \vec{r})
\end{pmatrix} \, , 
\end{eqnarray}
\end{widetext}
where $S_{12}$ is defined as
\begin{eqnarray}
S_{12}(\vec{a}_1, \vec{b}_2, \hat{r}) &=&
3\,\vec{a}_1 \cdot \hat{r}\,\vec{b}_2 \cdot \hat{r} - \vec{a}_1 \cdot \vec{b}_2
\, .
\end{eqnarray}

\subsection{Partial Wave Projection}

We can simplify the representation of the OPE potential by projecting
into ${\rm H \bar H}$ states with well-defined $J^{PC}$
quantum numbers.
This tasks is greatly simplified in coordinate space, for which only
the operators ${\bf C}_{12}(\hat{r})$ and ${\bf S}_{12}(\hat{r})$
are to be projected.
In the $\rm P\bar{P}$ case, we consider the partial waves
\begin{eqnarray}
| P\bar{P} (ljm) \rangle &=& \delta_{jl} \, {Y}_{lm}(\hat{r}) \, ,
\end{eqnarray}
where ${Y}_{lm}(\hat{r})$ is a spherical harmonic.
The parity and C-parity of this partial wave are $P = C = (-1)^l$.
The $\rm P\bar{P}^*$ / $\rm P^*\bar{P}$ cases are a bit more involved
owing to the polarization of the $\rm P^{*}$/$\rm \bar P^{*}$ vector meson.
The partial waves with well-defined orbital and total angular momentum $l$
and $j$ are defined as
\begin{eqnarray}
| P \bar{P}^* (l j m) \rangle &=& \sum_{m_l, \nu}
Y_{l m_l}(\hat{r}) | 1 \nu \rangle \langle l m_l 1 \nu | j  m \rangle \, , \\
| P^* \bar{P} (l j m) \rangle &=& \sum_{m_l, \mu}
Y_{l m_l}(\hat{r}) | 1 \mu \rangle \langle l m_l 1 \mu | j  m \rangle \, ,
\end{eqnarray}
where $| 1 \mu \rangle$, $| 1 \nu \rangle$ are the polarization vectors
of particle $1$ and $2$ and $\langle l_1 m_1 l_2 m_2 | j  m \rangle$
a Clebsch-Gordan coefficient.
The polarization vectors are to be understood as giving matrix elements
of the type
\begin{eqnarray}
\langle 1 \mu' | {\vec{\epsilon}_{1'}\,}^{*} \cdot \vec{\epsilon}_1
| 1 \mu \rangle &=& {\hat{e}_{\mu'}}^{*} \cdot \hat{e}_{\mu} \, , \\
\langle 1 \nu' | {\vec{\epsilon}_{2'}\,}^{*} \cdot \vec{\epsilon}_1
| 1 \mu \rangle &=& {\hat{e}_{\nu'}}^{*} \cdot \hat{e}_{\mu} \, ,
\end{eqnarray}
and so on,
where we have added primas to denote final states and with
$\hat{e}_{\mu}$ the unit vector in the spherical basis.
A problem with the two partial waves we have written above is that
they do not have a well-defined C-parity.
Thus, we define the linear combination
\begin{eqnarray}
| P \bar{P}^*(\eta) (l j m) \rangle = \frac{1}{\sqrt{2}}\,\left[
| P \bar{P}^* (l j m) \rangle  - \eta | P^* \bar{P} (l j m)\rangle \right] \, ,
\nonumber \\
\end{eqnarray}
with intrinsic C-parity $\eta$ and total C-parity $C = (-1)^l\,\eta$.
Finally, the $\rm P^*\bar{P}^*$ partial wave with intrinsic, orbital and
total angular momentum $s$, $l$ and $j$ is given by
\begin{eqnarray}
| P^* \bar{P}^* (s l j m) \rangle &=& \sum_{m_l, m_s}
Y_{l m_l}(\hat{r}) | s m_s \rangle \langle l m_l s m_s | j  m \rangle \, ,
\nonumber \\
\end{eqnarray}
where the spin state $| s m_s \rangle$ is
\begin{eqnarray}
| s m_s \rangle = \sum_{\mu, \nu}
| 1 \mu \rangle | 1 \nu \rangle \langle 1 \mu 1 \nu | s m_s \rangle \, .
\end{eqnarray}
The C-parity of these states is $C = (-1)^{l+s}$.
Alternatively we can also use the spectroscopic notation $^{2s+1}l_j$
instead of $(s)lj$ for denoting the partial waves.

From the previous definitions we can compute the matrix elements of
the central and tensor operators as
\begin{eqnarray}
\langle (s') l' j' m' | \hat{\bf O}_{12} | (s) l j m \rangle &=&
\nonumber \\
\int d^2 \hat{r}\,
\langle (s') l' j' m' | \hat{\bf O}_{12}(\hat{r}) | (s) l j m \rangle 
&=& \nonumber \\
\delta_{j j'} \delta_{m m'} \, {\bf O}^j_{(s' s)\, l' l} \, ,
\end{eqnarray}
where $\hat{\bf O}_{12}(\hat{r})$ either represents
$\hat{\bf C}_{12}(\hat{r})$ or $\hat{\bf S}_{12}(\hat{r})$.
The particle channels have been made implicit owing to the matrix notation.
The central and tensor operators preserve the $J^{PC}$ quantum numbers,
which for the particular case of the tensor operator means that the
orbital angular momentum can only change by an even number of units.
In addition, if we only consider the ${\rm P^* \bar P^*}$ particle channel,
the intrinsic angular momentum is also subjected to the restriction 
of even $| s - s' |$.
However, terms mixing the ${\rm P^* \bar P^*}$ and 
${\rm P \bar P^*}$/${\rm P^* \bar P}$ channels
can change $s$ by one unit. 

Thus  the most compact way to write the matrix elements of
the $\hat{\bf C}_{12}(\hat{r})$ or $\hat{\bf S}_{12}(\hat{r})$ operators
is to consider a particle/partial wave basis
with well-defined $J^{PC}$.
Examples are the following
\begin{eqnarray}
\mathcal{B}_{\rm H \bar H}(0^{++}) &=& \Big\{ | ^1S_0 (P\bar{P}) \rangle,
| ^1S_0 (P^*\bar{P}^*) \rangle, \nonumber \\
&& \phantom{ \Big\{ }  | ^5D_0 (P^*\bar{P}^*) \rangle \Big\} \, , \\
\mathcal{B}_{\rm H \bar H}(1^{+-}) &=& \Big\{
| ^3S_1 (P\bar{P}^* + P^*\bar{P}) \rangle, \nonumber \\ && \phantom{ \Big\{ }
| ^3D_1 (P\bar{P}^* + P^*\bar{P}) \rangle, \nonumber \\ && \phantom{ \Big\{ }
| ^3S_1 (P^*\bar{P}^*) \rangle,
| ^3D_1 (P^*\bar{P}^*) \rangle \Big\} \, , 
\end{eqnarray}
\begin{eqnarray}
\mathcal{B}_{\rm H \bar H}(2^{++}) &=& \Big\{
| ^1D_2 (P\bar{P}) \rangle,
| ^3D_2 (P\bar{P}^* - P^*\bar{P}) \rangle, \nonumber \\ && \phantom{ \Big\{ }
| ^1D_2 (P^*\bar{P}^*) \rangle,
| ^5S_2 (P^*\bar{P}^*) \rangle, \nonumber \\ && \phantom{ \Big\{ }
| ^5D_2 (P^*\bar{P}^*) \rangle,
| ^5F_2 (P^*\bar{P}^*) \rangle
\Big\} \, , 
\end{eqnarray}
for which we obtain the central matrices
\begin{eqnarray}
{\bf C}_{12}(0^{++}) &=& 
\begin{pmatrix}
0 & \sqrt{3} & 0 \\
\sqrt{3} & -2 & 0 \\
0 & 0 & 1
\end{pmatrix} \, , 
\end{eqnarray}
\begin{eqnarray}
{\bf C}_{12}(1^{+-}) &=& 
\begin{pmatrix}
-1 &  0 & 2  & 0  \\
 0 & -1 & 0  & 2  \\
 2 & 0  & -1 & 0  \\
 0 & 2  &  0 & -1 \\
\end{pmatrix} \, ,
\end{eqnarray}
\begin{eqnarray}
{\bf C}_{12}(2^{++}) &=& 
\begin{pmatrix}
 0 &  0 & \sqrt{3}  & 0 & 0 & 0 \\
 0 &  1 & 0  & 0 & 0 & 0 \\
 \sqrt{3} &  0 & -2  & 0 & 0 & 0 \\
 0 &  0 & 0  & 1 & 0 & 0 \\
 0 &  0 & 0  & 0 & 1 & 0 \\
 0 &  0 & 0  & 0 & 0 & 1 \\
\end{pmatrix} \, ,
\end{eqnarray}
and the tensor matrices
\begin{eqnarray}
{\bf S}_{12}(0^{++}) &=& 
\begin{pmatrix}
0 & 0 & -\sqrt{6} \\
0 & 0 & -\sqrt{2} \\
-\sqrt{6} & -\sqrt{2} & -2
\end{pmatrix} \, , \\
{\bf S}_{12}(1^{+-}) &=& 
\begin{pmatrix}
0 & \sqrt{2} & 0 & \sqrt{2} \\
\sqrt{2} & -1 & \sqrt{2} & -1 \\
0 & \sqrt{2} & 0 & \sqrt{2} \\
\sqrt{2} & -1 & \sqrt{2} & -1 \\
\end{pmatrix} \, ,
\end{eqnarray}
\begin{widetext}
\begin{eqnarray}
{\bf S}_{12}(2^{++}) &=&
\begin{pmatrix}
0 & 0 & 0 & -\sqrt{\frac{6}{5}} & 2\sqrt{\frac{3}{7}} & -6\,\sqrt{\frac{3}{35}} \\
0 & -1 & 0 & 3\sqrt{\frac{2}{5}} & -\frac{3}{\sqrt{7}} & 
-\frac{12}{\sqrt{35}} \\
0 & 0 & 0 & -\sqrt{\frac{2}{5}} & \frac{2}{\sqrt{7}} & -\frac{6}{\sqrt{35}} \\
-\sqrt{\frac{6}{5}} & 3\sqrt{\frac{2}{5}} & -\sqrt{\frac{2}{5}} & 0 & \sqrt{\frac{14}{5}} & 0 \\
2\sqrt{\frac{3}{7}} & -\frac{3}{\sqrt{7}} & \frac{2}{\sqrt{7}} & \sqrt{\frac{14}{5}} & \frac{3}{7} & 
\frac{12}{7\sqrt{5}} \\
-6\,\sqrt{\frac{3}{35}} & \frac{12}{\sqrt{35}}  & -\frac{6}{\sqrt{35}} & 0 & \frac{12}{7\sqrt{5}} & -\frac{10}{7}
\end{pmatrix} \, .
\end{eqnarray}
\end{widetext}
Other $J^{PC}$ possibilities are to be computed as these three examples.
As can be seen, the resulting coupled channel structure can be quite complex:
in certain cases we can have up to six coupled channels.
The simplifying feature is that particle coupled channel effects are suppressed
by two orders in the chiral expansion of the potential,
which means that in low order calculations
we can concentrate in a specific particle channel and ignore the others.
In this case, we only need to consider the diagonal terms of the central
operator and, in what regard the tensor operator, we end up with
the matrices shown from Eq.~(\ref{eq:S12_PPst_3UJ}) to
(\ref{eq:S12_PstPst_15CJ}) with at most four coupled channels.


%

\end{document}